\documentclass[12pt]{article}
\usepackage{amsmath,amsfonts}
\usepackage{hyperref}
\usepackage{cancel}
\usepackage[normalem]{ulem}
\usepackage{color}

\numberwithin{equation}{section}

\definecolor{blue-violet}{rgb}{0.54, 0.17, 0.89}
\definecolor{PineGreen}{cmyk}{0.92, 0, 0.59, 0.25}
\definecolor{OliveGreen}{cmyk}{0.64, 0, 0.95, 0.40}
\definecolor{RawSienna}{cmyk}{0, 0.72, 1, 0.45}
\definecolor{Gray}{cmyk}{0, 0, 0, 0.50}
\definecolor{MidnightBlue}{cmyk}{0.98, 0.13, 0, 0.43}
\definecolor{Orange}{cmyk}{0, 0.61, 0.87, 0}
\definecolor{LimeGreen}{cmyk}{0.50, 0, 1, 0}
\definecolor{Green}{cmyk}{1, 0, 1, 0}

\newcommand{\mc}{\textcolor{magenta}}

\begin{document}

\title{Energy of anti-de Sitter black holes in odd-dimensional Quadratic Curvature Gravity}

\author{Olivera Miskovic$^{1,}$\thanks{olivera.miskovic@pucv.cl} , Rodrigo Olea$^{2,}$\thanks{rodrigo.olea@unab.cl}\;  and Yoel Parra-Cisterna$^{1,}$\thanks{yoel-parra@live.cl} \bigskip\\
%{\small $^1$ Departamento de Física, Universidad de Buenos Aires y  IFIBA-CONICET,}\\
%{\small  Ciudad Universitaria, Pabellón 1, 1428 Buenos Aires, Argentina}\\
{\small $^1$ Instituto de F\'\i sica, Pontificia Universidad Cat\'olica de Valpara\'\i so,}\\
{\small Casilla 4059, Valpara\'\i so, Chile}\\
{\small $^2$  Departamento de Ciencias Físicas, Universidad Andres Bello,}\\
{\small Sazié 2212, Piso 7, Santiago, Chile }
} 

\maketitle
%\tableofcontents

\begin{abstract}
We provide new formulas for the energy of black hole solutions  in the anti-de Sitter (AdS) sector of a generic gravity theory with curvature-squared terms. This is achieved  by the addition of counterterms of the extrinsic type (\emph{Kounterterms}) which produces a finite variation of the total action. The procedure is compatible with the boundary conditions in asymptotically AdS spaces within the framework of Quadratic Curvature Gravity (QCG) theory. As a consequence, without resorting to background-substraction, or linearized methods, the conserved charges correctly reproduce the energy of static black holes in QCG.
\end{abstract}

\section{Introduction}
%
%
%\rc{Red `rc',} \bc{Blue `bc',} \cc{Cyan `cc',} \mc{Magenta `mc',} \vc{Violet `vc',}  \gc{Pine Green `gc',} \br{Brown `br',} \gr{Gray `gr',} \mb{Midnight Blue `mb',} \oc{Orange `oc',} \lc{Lime Green `lc'.}

The first attempts to apply quantization techniques to gravitational theory were made in the  1970's. By that time, it had already been possible to quantize the other fundamental forces as gauge theories. However, it was found that General Relativity shows divergences in the ultraviolet quantum regime  at the level of the propagator \cite{tHooft:1974toh,Goroff:1985th}. On the other hand, it was argued that the inclusion of quadratic terms in curvature into the action might lead to a renormalizable quantum theory of gravity \cite{Stelle:1976gc}.  

A few decades later, in the context of Holography, quadratic corrections of curvature regained interest in the community.  Clearly, by giving rise to fourth-order field equations, it enriches the asymptotic structure of spacetime. This richness is reflected in  the appearance of new  holographic sources at the conformal boundary, in addition to  new parameters to build a holograhic theory. It also allows for the emergence of solutions with a relaxed anti-de Sitter (AdS) asymptotic behavior, logarithmic modes \cite{Grumiller:2008qz, Alishahiha:2011yb} and Liftshitz type of metrics \cite{Cai:2009ac}.

From a classical standpoint, the problem of definition of energy in gravitational theories whose equations of motion contain higher derivative terms, both for asymptotically flat and asymptotically AdS solutions, has been extensively addressed in the literature.
%an important problem treated by many authors, among them is the contribution made by Deser and Tekin in ref. \cite{Deser:2002jk}, which found a general formula for conserved quantities in any dimension, via linearization of the equations of motion.

Along this line, in refs.\cite{Giribet:2018hck, Giribet:2020aks}, it was proposed a formula for gravitational energy in a fully covariant form in an arbitrary theory of gravity including quadratic-curvature terms in even bulk dimensions via the Noether-Wald method \cite{Iyer:1994ys}. More precisely, the prescription consists in adding a topological term to the action such that it renormalizes its variation. From a finite surface term in the variation of the total action, finite asymptotic charges are derived. This idea had already been consistently applied in second-derivative gravity theories, such as Einstein-AdS \cite{Aros:1999kt} and Einstein-Gauss-Bonnet-AdS gravity \cite{Kofinas:2006hr}.

In the present work, we construct conserved charges in the AdS sector of Quadratic Curvature Gravity (QCG) in any odd dimension. This is carried out by adding up to the bulk action a suitable boundary term, which renders its variation both well defined and finite. Exploiting the invariance under diffeomorphisms of the total (renormalized) action, we obtain a new formula  for the conserved quantities, associated to the boundary isometries. This expression produces the energy of static asymptotically AdS black holes in the theory, which matches the one found by other existing methods.  The generalization of the formula for the vacuum energy in the case of QCG turns this procedure interesting in the context of Holography.

%%%%%%%%%%%%%%%%%%%%%%%%%%%%%%%%%%%%%%%%%%%%%%%%%%%%%%
\section{Kounterterms in AdS gravity} \label{K in AdS}

Gauge/gravity duality realized in the form of anti-de Sitter/Conformal Field Theory (AdS/CFT) correspondence can be made explicit through the semi-classical  relation 
\begin{equation}\label{Partition}
    Z_{\mathrm{CFT}}[\phi _{0}]\approx \left. \exp (iI_{\mathrm{grav}}[\phi ])\right\vert
_{\phi \rightarrow \phi _{0}}\,,
\end{equation}
for bulk fields $\phi $ approaching the value $\phi _{0}$ at the conformal
boundary of the spacetime.
The above formula equals the partition function of the $d$-dimensional CFT to the exponential of the classical $(d+1)$-dimensional gravity action \cite{Maldacena:1997re, Gubser:2002tv, Witten:1998qj}.

For a spacetime metric written in normal coordinates, the line element can be always cast in the form
\begin{equation}\label{NormalCoordinates}
    ds^{2}= g_{\mu \nu }(x)\, dx^{\mu }dx^{\nu }=N(z)^{2}%
\,dz^{2}+h_{ij}(x,z)\,dx^{i}dx^{j}\,.
\end{equation}
In the case of pure AdS gravity, the bulk metric can be always written down in Fefferman-Graham (FG) form with
\begin{equation}\label{defN}
    N(z)=\frac{\ell}{z}\,,\hspace{1.5cm}
h_{ij}(z,x)=\frac{1}{z^{2}}\,\tilde{g}_{ij}(x,z)\,,
\end{equation}
 which adequately represents  a pole of order two at the conformal boundary (located at the radial position $z=0$, where $\tilde{g}_{ij}$ is regular). This is a defining feature of any asymptotically AdS (AAdS) spacetime, as long as the metric $\tilde g_{ij}(x,z)$ admits a series expansion of the form
\begin{equation}\label{FGexpansion}
    \tilde g_{ij} (x,z)=g_{(0)ij}(x)+z^{2}g_{(2)ij}(x)+z^{4}g_{(4)ij}(x)+\cdots ,
\end{equation}
where only even powers of the radial coordinate appear until the holographic order $\mathcal{O}(z^{d})$. 

The asymptotic resolution of the field equations leads to the coefficients $g_{(k)ij}$ as functionals of the holographic metric $g_{(0)ij}$. This method defines Holographic Renormalization procedure, which enables to identify and remove divergent contributions as negative powers in $z$ both in the action and its corresponding variation \cite{Skenderis:2002wp, Henningson:1998gx}.

In turn, this implies the addition of local counterterms to the bulk gravity action, on top of the Gibbons-Hawking term
\begin{equation}\label{EH+GH+Ct}
I_\mathrm{ren}=\dfrac{1}{16\pi G}\int\limits_{M}d^{d+1}x\,\sqrt{-g}
\,\left( R-2\Lambda \right) -\dfrac{1}{8\pi G}\int\limits_{\partial
M}d^{d}x\,\sqrt{-h}\,K+\int\limits_{\partial M}d^{d}x\,L_{ct}(h,\mathcal{R},D\mathcal{R})\,,
\end{equation}
where the cosmological constant is $\Lambda =-\frac{d(d-1)}{2\ell ^{2}}$, in terms of the AdS radius $\ell $. In the above relation, the boundary metric is $h_{ij}= \tilde g_{ij}(x,z)/z^{2}$, $K$ is the trace of the extrinsic curvature $K_{ij}$ defined in the frame (\ref{NormalCoordinates}) as
\begin{equation}\label{Kextrinsic}
    K_{ij}=-\frac{1}{2}n^{\mu }\partial _{\mu }h_{ij}\,,
\end{equation}
in terms of the normal to the boundary $n_{\mu }=N\delta _{\mu }^{z}$.

 The series of counterterms $L_{ct}$  is constructed with intrinsic tensors defined with the boundary metric, namely its curvature $\mathcal{R}^{i}_{\ jkl}(h)$ and covariant derivatives thereof. Naively speaking, the addition of counterterms of a different sort would be at odds with a variational principle consistent with a Dirichlet boundary condition for $h_{ij}$. 

This allows to obtain a renormalized version of the Brown-York stress tensor
\begin{equation}
T_\mathrm{ren}^{ij}[h]=\pi ^{ij}+\frac{2}{\sqrt{-h}}\,\frac{\delta L_{ct}}{\delta h_{ij}}\,,  \label{Tren}
\end{equation}
in terms of the canonical momentum
\begin{equation}
\pi ^{ij}=\frac{1}{8\pi G}\left( K^{ij}-h^{ij}K\right) \,,  \label{pi}
\end{equation}
associated to the radial evolution of the spacetime geometry.

Interestingly enough, holographic properties of the stress tensor (e.g., Weyl anomaly) can be derived from a suitable rescaling of the quasilocal energy-momentum tensor 
\begin{equation}
    T^{ij}[g_{(0)}]=\lim\limits_{z\rightarrow 0}\left( \frac{1}{z^{d-2}}\, T^{ij}[h]\right) \,.
\end{equation}
This is a well-defined, systematic algorithm for the asymptotic resolution of the field equations in terms of the conformal data. However, it turns increasingly complicated in higher dimensions, let alone in higher-curvature gravity theories. In practice, one lacks a general prescription for the counterterm series in all such situations.

It is therefore of utmost importance to explore a different pathway to renormalization of AdS gravity and holographic correlators derived from it.  There are several arguments which lead to the idea that such a thing is plausible.

The first one is the fact that a Dirichlet action principle for the metric $\delta h_{ij}|_{\partial M}=0$  is ill defined in AdS gravity. Indeed, in the FG frame,
the expansion of the metric is 
\begin{equation}
    h_{ij}=\frac{g_{(0)ij}}{z^{2}}+\cdots\,,
\end{equation}
which blows up at the conformal boundary. A consistent way out is to demand instead a variational problem for the metric $g_{(0)ij}$ at the conformal boundary, which is compatible with the holographic description of AdS gravity. On the other hand,  this infrared behavior of the metric will induce divergences in the variation of the action which will require counterterms to be cured. In that respect, this reasoning shows the link between the variational principle associated to $g_{(0)ij}$ and the renormalization problem \cite{Papadimitriou:2005ii}.

At the same time, there is a blissful accident in AdS gravity: the
leading-order term of the extrinsic curvature is the same as the one in the boundary metric (up to a proper length scale). As a matter of fact, for asymptotically AdS spaces, the extrinsic curvature adopts the form \cite{Witten:2018lgb}
\begin{equation}
    K_{ij}=\frac{1}{\ell }\frac{g_{(0)ij}}{z^{2}}+\cdots\,,
\end{equation}
such that it makes manifest that the above claim is not longer true in asymptotically flat gravity ($\ell \rightarrow \infty $). This simple observation opens the possibility of adding up counterterms to the action, which may depend also on the extrinsic curvature
\begin{equation}
    \tilde{I}_{\mathrm{ren}}=I_{\mathrm{bulk}}+c_{d}\int\limits_{\partial M}d^{d}x\,\sqrt{-h}\, B_{d}(f(h),K)\,,
\end{equation}
as long as the boundary description of the theory turns holographic, i.e., the variation of the total action is finite and expressible in terms of the
variation of the holographic source\footnote{This far, one cannot guarantee that $\tilde{I}_{\mathrm{ren}}$ would coincide with the standard renormalized action. The same applies to the holographic
correlator $\tau ^{ij}$.}
\begin{equation}
    \delta \tilde{I}_{\mathrm{ren}}=\frac{1}{2}\int\limits_{\partial M}d^dx\,\sqrt{-g_{(0)}}\, \tau ^{ij}\,\delta g_{(0)ij}\,.
\end{equation}
An additional interesting observation comes from one of the simplest definition of conserved charges in AdS gravity, which is Conformal Mass \cite{Ashtekar:1984zz,Ashtekar:1999jx}. For a given set of boundary Killing vectors $\{\xi ^{i}\}$, a conserved quantity can be derived as a surface integral
\begin{equation}
    Q_{\mathrm{AMD}}[\xi ]=\frac{1}{8\pi G}\frac{\ell }{d-2} \int\limits_{\Sigma }d^{d-1}x\sqrt{\sigma}\,u_{i}\,\mathcal{E}_{j}^{i}~\xi ^{j}\,,
\end{equation}
in a co-dimension two region $\Sigma $ with the (induced) metric $\sigma_{mn}$ and the timelike unit normal $u_i$. 
AMD stands for Ashtekar-Magnon-Das. Here, the tensor $\mathcal{E}_{j}^{i} $ is the electric part of the Weyl tensor, obtained as the double contraction with a spacelike normal vector to the boundary $n_{\mu }$, i.e.,
\begin{equation}
    \mathcal{E}_{j}^{i}=W_{\ \ j\nu }^{i\beta }\,n_{\beta }n^{\nu }\,,
\end{equation}
or, equivalently, in the frame \eqref{NormalCoordinates},
\begin{equation}
\mathcal{E}_{j}^{i}=W_{\ \ jz}^{iz}=-W_{\ \ jk}^{ik}\,, \label{EasWtrace}
\end{equation}
in terms of the boundary subtrace.
AMD charges correctly reproduce the mass and angular momentum of AAdS black holes. However, in contrast to the quasilocal stress tensor, the electric part of Weyl tensor is traceless ($\mathcal{E}_{i}^{i}=0$), while the trace of the stress tensor features an anomaly at a holographic order
\begin{equation}
    T_{i}^{i}=z^{d}\mathcal{A\,}\,,
\end{equation}
for $d$ even. Therefore, one expects that, in general, $T_{j}^{i}$ may
differ from $\mathcal{E}_{j}^{i}$ by subleading, holographic contributions $\Delta _{j}^{i}$, whose trace reproduces the conformal anomaly. As a matter of fact, this was explicitly checked for $d+1=5$ dimensions \cite{Ashtekar:1999jx},
such that
\begin{equation}
    T_{j}^{i}=\mathcal{E}_{j}^{i}+\Delta _{j}^{i}\,,\quad \Delta _{i}^{i}=z^{4} \mathcal{A\,}.
\end{equation}
Furthermore, for Einstein gravity, the Weyl tensor takes the particular form
\begin{equation}
    W_{\mu \nu }^{\alpha \beta }=R_{\mu \nu }^{\alpha \beta }+\frac{1}{\ell ^{2}}\,\delta _{\mu \nu}^{\alpha \beta}\,,
\end{equation}
such that, its electric part can be expressed as 
\begin{equation}
    \mathcal{E}_{j}^{i}=-\left( R_{\quad jk}^{ik}+\frac{d-1}{\ell ^{2}}\,\delta _{j}^{i}\right) \,.
\end{equation}

Therefore, using Gauss-Codazzi relations in five spacetime dimensions, the stress tensor becomes
\begin{equation}
    T_{j}^{i}=-\left( \mathcal{R}_{j}^{i}-K_{j}^{i}K+K_{k}^{i}K_{j}^{k}+\frac{3}{%
\ell ^{2}}\delta _{j}^{i}\right) +\Delta _{j}^{i}\,,
\end{equation}
from which it is evident that the stress tensor contains quadratic terms in the extrinsic curvature. This may look puzzling at first glance when compared to eqs.(\ref{Tren}) and (\ref{pi}), as the stress tensor is at most linear in $K_{ij}$ plus purely intrinsic contributions. Then, one understands that the asymptotic expansion of the fields in AdS gravity is such that this last formula can be truncated to the linear order in $K_{ij}$ without loss of holographic information. This can be made, in a consistent way, using FG frame and a power-counting argument.

The moral of the above point is the fact one may equally consider boundary terms to renormalize AdS gravity action which are nonlinear in the extrinsic curvature (Kounterterms). These terms should be the resummation of the counterterm series into a compact polynomial of the extrinsic and intrinsic curvatures.

With the definition of an auxiliary tensor with the symmetries of the Riemann tensor,
\begin{equation}
 \mathcal{F}_{j\,l}^{i\,k}(x,y)=\frac{1}{2}\,\left(\mathcal{R}_{j\,l}^{i\,k}-x^{2}(K_{j}^{i}K_{l}^{k}-K_{l}^{i}K_{j}^{k})+\frac{y^{2}}{\ell^2}\,\delta _{jl}^{ik}\right)\,,    
\end{equation}
in even-dimensional AAdS gravity ($d+1=2n$), the corresponding Kounterterms take a compact form, as a result of the use of a parametric integral  \cite{Olea:2005gb}
%\begin{eqnarray}
%B_{2n-1} &=&2n\int\limits_{0}^{1}du\,\delta _{i_{1}\cdots
%i_{2n-1}}^{j_{1}\cdots j_{2n-1}}\,K_{j_{1}}^{i_{1}}\left( \frac{1}{2}\,\mathcal{R}_{j_{2}\,j_{3}}^{i_{2}\,i_{3}}-u^{2}K_{j_{2}}^{i_{2}}K_{j_{3}}^{i_{3}}\right) \times \cdots
%\notag \\
%&&\cdots \times \left( \frac{1}{2}\,\mathcal{R}_{j_{2n-2}j_{2n-1}}^{i_{2n-2}i_{2n-1}}-u^{2}K_{j_{2n-2}}^{i_{2n-2}}K_{j_{2n-1}}^{i_{2n-1}}\right) \,.
%\label{EvenKounterterms}
%\end{eqnarray}
%
\begin{equation}
B_{2n-1} =2n\int\limits_{0}^{1}du\,\delta _{i_{1}\cdots
i_{2n-1}}^{j_{1}\cdots j_{2n-1}}\,K_{j_{1}}^{i_{1}}\,\mathcal{F}_{j_{2}\,j_{3}}^{i_{2}\,i_{3}}(u,0) \cdots 
\mathcal{F}_{j_{2n-2}j_{2n-1}}^{i_{2n-2}i_{2n-1}}(u,0) \,.
\label{EvenKounterterms}
\end{equation}
 In this case, the Kounterterms are proportional to the $n$-th Chern form. By virtue of the Euler theorem, they can be well traded off by the corresponding Euler term in the bulk, up to the Euler characteristic of the manifold.\footnote{Conventions and properties of the Kronecker symbol of order $p$, as well as useful integrals to manipulate surface terms, are given in Appendix \ref{Conventions}.}

As for odd spacetime dimensions, the Kounterterms adopt a slightly different form, which is a consequence of a double-parametric integral \cite{Olea:2006vd}
%\begin{eqnarray}
%B_{2n} &=&2n\int\limits_{0}^{1}du\int\limits_{0}^{u}ds\,\delta
%_{i_{1}\cdots i_{2n}}^{j_{1}\cdots j_{2n}}K_{j_{1}}^{i_{1}}\delta
%_{j_{2}}^{i_{2}}\left(\frac{1}{2}\,\mathcal{R}_{j_{3}\,j_{4}}^{i_{3}\,i_{4}}-u^{2}K_{j_{3}}^{i_{3}}K_{j_{4}}^{i_{4}}+\frac{s^{2}}{\ell^2}\, \delta _{j_{3}}^{i_{3}}\delta _{j_{4}}^{i_{4}}\right) \times \cdots \notag \\
%&&\qquad \cdots \times \left( \frac{1}{2}\,\mathcal{R}_{j_{2n-1}%
%\,j_{2n}}^{i_{2n-1}\,i_{2n}}-u^{2}K_{j_{2n-1}}^{i_{2n-1}}K_{j_{2n}}^{i_{2n}}+\frac{s^{2}}{\ell^2}\,\delta _{j_{2n-1}}^{i_{2n-1}}\delta_{j_{2n}}^{i_{2n}}\right) \,.  %\label{KountertermOdd}
%\end{eqnarray}

\begin{equation}
B_{2n} =2n\int\limits_{0}^{1}du\int\limits_{0}^{u}ds\,\delta
_{i_{1}\cdots i_{2n}}^{j_{1}\cdots j_{2n}}K_{j_{1}}^{i_{1}}\delta
_{j_{2}}^{i_{2}}\,\mathcal{F}_{j_{3}\,j_{4}}^{i_{3}\,i_{4}}(u,s) \cdots
 \,\mathcal{F}_{j_{2n-1}\,j_{2n}}^{i_{2n-1}\,i_{2n}}(u,s) \,.  
 \label{KountertermOdd}
\end{equation}

The last formula is not a standard structure in the mathematical toolbox. Notwithstanding the foregoing,  there are striking similarities between this object and the Chern form. Indeed,  both $B_{2n-1}$ and $B_{2n}$, when consistently added to the Euclidean bulk action, are able to reproduce the correct thermodynamics for AAdS black holes. Furthermore, in the holographic framework, they are able to renormalize AdS gravity action for AAdS space with conformally flat boundary \cite{Miskovic:2009bm, Anastasiou:2020zwc}\footnote{A recent proposal to circumvent this limitation considers the proper embedding of Einstein-AdS theory in Conformal gravity \cite{Anastasiou:2020mik}.}. In addition, when evaluated on manifolds with a conical defect, the singular part of $B_{2n}$ is $B_{2n-2}$, a self-replication in two dimensions lower, what is also a property of the Chern form \cite{Anastasiou:2019ldc}.

In the next section, we review the main properties, as regards bulk dynamics and surface terms, of a gravity theory that includes arbitrary quadratic couplings in the curvature.

%%%%%%%%%%%%%%%%%%%%%%%%%%%%%%%%%%%%%%%%%%%%%%%%%%%%%%
\section{Quadratic Curvature Gravity}
\label{SecQCG}

Quadratic Curvature Gravity (QCG) is obtained from adding to Einstein-Hilbert's action all the possible quadratic terms of the curvature, that is 
\begin{equation}\label{QCGaction}
    I_{\mathrm{QCG}}=\int\limits_{M}\!\!d^{d+1}x\sqrt{-g}\left[\frac{1}{\kappa}\,(R-2\Lambda_{0}) +\alpha R_{\mu \nu}R^{\mu \nu}+\beta R^{2}+\gamma\, GB\right],
\end{equation}
which are of fourth order in the derivatives. Here, $\kappa=2 \mathrm{Vol}(S^{d-1}) G$ is the gravitational constant and  $\Lambda_0$ is the bare cosmological constant. 

In this action, the Riemann-squared term  $R_{\mu\nu\alpha\beta}R^{\mu\nu\alpha\beta}$ has been traded off by the Gauss-Bonnet, $R_{\mu\nu\alpha\beta}R^{\mu\nu\alpha\beta}-4R_{\mu\nu}R^{\mu\nu}+R^2$. This has been denoted by $GB$, which does not produce higher-derivative terms in the metric. Thus, higher-derivative terms are associated only to the couplings $\alpha$ and $\beta$.
The action \eqref{QCGaction} reduces to Einstein-Gauss-Bonnet gravity with a cosmological constant when $\alpha=0$ and $\beta=0$. Natural dimensions of the coupling constants appearing in the action are $[\kappa]=L^{d-1}$ and $[\alpha]=[\beta]=[\gamma]=1/L^{d-3}$.

The variation of the action (\ref{QCGaction}) with respect to the metric  $g_{\mu\nu}(x)$ leads to the equation of motion
\begin{equation}
E_{\mu \nu }=\frac{1}{\kappa }\,G_{\mu \nu }+\gamma H_{\mu \nu }+P_{\mu \nu}=0\,,  \label{eomQCG}
\end{equation}
with the Einstein's tensor
\begin{equation}
 G_{\mu \nu}=R_{\mu \nu}-\frac{1}{2}\,g_{\mu \nu}\,R+\Lambda_{0} \, g_{\mu \nu}  \,,
\end{equation}
and $H_{\mu \nu}$ is the Lanczos tensor, 
\begin{eqnarray}\label{Lanczos}
H_{\mu\nu} &=& -\frac{1}{2}\,g _{\mu \nu }\left( R^{2}-4R^{\alpha \beta }R_{\alpha \beta }+R^{\alpha \beta \lambda \sigma }R_{\alpha \beta \lambda \sigma }\right) \notag\\
&&+2\left( RR_{\mu \nu }-2R_{\mu \lambda }R^{\lambda}_{\ \nu }-2R_{\mu \alpha \nu \beta }R^{\alpha \beta }+R_{\mu \lambda \alpha \beta }R_{\nu } ^{\ \lambda \alpha \beta }\right) \,,  
\end{eqnarray}
which describes the contribution of the Gauss-Bonnet term to the equations of motion.
In dimension $d+1=4$, $H_{\mu \nu}$ vanishes identically because it comes from a topological term.

The contribution of the Ricci scalar-squared and Ricci tensor-squared terms to the equations of motion is given by the symmetric tensor
\begin{eqnarray}
P_{\mu \nu } &=&2\beta R\left( R_{\mu \nu }-\frac{1}{4}\,g_{\mu \nu
}R\right) +\left( \alpha +2\beta \right) \left( g_{\mu \nu }\square -\nabla
_{\mu }\nabla _{\nu }\right) R  \notag \\
&&+\alpha \square G_{\mu \nu }+2\alpha \left( R_{\mu \sigma \nu \lambda }-
\frac{1}{4}\,g_{\mu \nu }R_{\sigma \lambda }\right) R^{\sigma \lambda }\,.
\end{eqnarray}
The field equations of QCG are fourth order differential equations, i.e., quadratic terms in the curvature or second covariant derivatives of them  in the tensor $P_{\mu\nu}$. 

The presence of new dimensionful couplings, in principle, may modify the bare cosmological constant $\Lambda _{0}$ into an effective one, $\Lambda _{\mathrm{eff}}$. While the theory features both Einstein and non-Einstein solutions, it is reasonable to assume that both branches share a common vacuum state, defined as global constant-curvature spaces
\begin{equation}
R_{\alpha \beta }^{\mu \nu }-\frac{2\Lambda _{\mathrm{eff}}}{d(d-1)}\,\delta _{\alpha \beta }^{\mu \nu }=0\,.  \label{RiemannQCG}
\end{equation}

These maximally symmetric spaces correspond to the vacuum states of the theory. Therefore, massive solutions correspond to deformations on the geometry of pure (A)dS space, such that the right-hand side of the above relation is proportional to the mass of the object under study.

Plugging in the condition (\ref{RiemannQCG}) in the equation of motion (\ref{eomQCG}), we find
\begin{eqnarray}
G_{\mu \nu } &=&\left( \Lambda _{0} -\Lambda_{\mathrm{eff}}\right) \,g_{\mu\nu }\,,  \qquad  H_{\mu \nu } =-\frac{2(d-2)(d-3) }{d(d-1)}\,\Lambda _{\mathrm{eff}}^{2}\,g_{\mu \nu }\,, \nonumber\\
P_{\mu \nu } &=&-\frac{2(d-3) }{(d-1)^{2}}\,\left( \alpha +(d+1)\beta\right) \,\Lambda_{\mathrm{eff}}^{2}\, g_{\mu \nu }\,.\label{Lambda_eff}
\end{eqnarray}

From the above relations, one obtains the effective cosmological constant in terms of the bare one, in the form \footnote{As seen from (\ref{Lambda_eff}), $\Lambda _{\mathrm{eff}}=0$ is possible only when $\Lambda _{0}=0$. We are not interested in this case.}
\begin{equation}\label{comparacionLambda}
    \frac{\Lambda_{0}}{2\kappa\Lambda^{2}_{\mathrm{eff}}}-\frac{1}{2\kappa\Lambda_{\mathrm{eff}}}=\frac{(d-3)}{(d-1)^{2}}\,\left(\alpha+(d+1)\beta\right)+\gamma\,\frac{(d-2)(d-3)}{d(d-1)}\,.
\end{equation}
This equation produces $\Lambda _{\mathrm{eff}}\neq \Lambda _{0}$ only when $d+1>4$. Indeed, $\Lambda _{\mathrm{eff}}=\Lambda _{0}$ is obtained either in four spacetime dimensions, or in higher dimensions when the coupling constants satisfy $\frac{\alpha +(d+1)\beta }{d-1}+\gamma \,\frac{d-2}{d}=0$, because then the r.h.s. in (\ref{comparacionLambda}) vanishes identically. In this case, as in Einstein-Hilbert gravity, the vacuum state of the system is unique as its effective cosmological constant is not altered by the addition of quadratic-curvature terms.

In dimensions higher than four, with $\Lambda _{\mathrm{eff}}$, $\Lambda_{0}\neq0$, the equation (\ref{comparacionLambda}) has two solutions, as it is a second order algebraic equation for $\Lambda_{\mathrm{eff}}$. The solutions of the quadratic equation have the form \footnote{In $d+1>4$, when $\Lambda _{0}=0$, it is also possible to have $\Lambda _{\mathrm{eff}}\neq 0$ of any sign. For the sake of simplicity, we will omit this particular case from our discussion, and focus only on a theory with generic values of $\alpha $, $\beta $ and $\gamma $.} 
\begin{equation}
    \frac{1}{\Lambda_{\mathrm{eff}}^{\pm}}=\frac{1}{2\Lambda_{0}}\left[ 1\pm\sqrt{1+8\kappa \Lambda _{0}\,\frac{d-3}{d-1}\left( \frac{\alpha +(d+1)\beta }{d-1}+\gamma \,\frac{d-2}{d}\right)}\right]\,.
\end{equation}
These solutions exist only if the constants $\alpha$, $\beta$ and $\gamma$ are such that
\begin{equation}\label{desigualdad}
    1+8\kappa \Lambda _{0}\,\frac{d-3}{d-1}\left( \frac{\alpha +(d+1)\beta }{d-1}+\gamma \,\frac{d-2}{d}\right)\geq0\,.
\end{equation}

The vacuum state is twofold degenerate in $d+1>4$, i.e., $\Lambda_{\mathrm{eff}}^{+}=\Lambda_{\mathrm{eff}}^{-}=2\Lambda_{0}$,  when the coupling constants are such that  (\ref{desigualdad}) vanishes. 
We will focus on AAdS spacetimes, defined in terms of the effective cosmological constant. For that case $\Lambda_{\mathrm{eff}}$ is related to the effective AdS radius $\ell_{\mathrm{eff}}$  as
\begin{equation}
    \Lambda_{\mathrm{eff}}=
    -\frac{d(d-1)}{2\ell_{\mathrm{eff}}^{2}}\,.
\end{equation}
Therefore, we will assume that QCG possesses two inequivalent asymptotically AdS branches in $d+1>4$, what is true for certain intervals of the parameters. When the coupling constants are small, the quadratic curvature terms can be treated as a small correction to the Einstein-AdS action. In that case, the expression (\ref{comparacionLambda}) can be expanded as
\begin{eqnarray}
\frac{1}{\Lambda _{\mathrm{eff}}^{+}} &=&\frac{1}{\Lambda _{0}}+2\kappa \,\frac{d-3}{d-1}\left( \frac{\alpha +(d+1)\beta }{d-1}+\gamma \,\frac{d-2}{d}\right) +\cdots \ ,  \notag \\
\frac{1}{\Lambda _{\mathrm{eff}}^{-}} &=&-2\kappa \,\frac{d-3}{d-1}\left(\frac{\alpha +(d+1)\beta }{d-1}+\gamma \,\frac{d-2}{d}\right) +\cdots \ .
\end{eqnarray}

It is thus clear that, in the \textit{Einstein} branch with $\Lambda_{\mathrm{eff}}^{+}$  we can recover Einstein solutions in the weak coupling limit. On the contrary, the \textit{stringy} branch with $\Lambda_{\mathrm{eff}}^{-}$, is not continuously connected to the Einstein theory \cite{Boulware:1985wk}.

The quadratic terms in the action also enrich the linearized form of the theory. By studying the propagating modes of the metric with respect to the AdS background, it is possible to find, in addition to the usual massless spin 2 mode, two new ones: a massive scalar mode, and a massive ghost-like spin-2 mode \cite{Bueno:2016ypa}.

 The surface term of the theory arises from integrating by parts the variation of the gravitational action in order to construct the equation of motion, that is,
 \begin{equation}
\delta I_{\mathrm{QCG}}=\int\limits_{M} d^{d+1}x\sqrt{-g}\, E_{\mu \nu}\delta g^{\mu \nu} +\int\limits_{M} d^{d+1}x\sqrt{-g}\,
\nabla _{\alpha }\Theta ^{\alpha}.
 \end{equation}
 For an arbitrary gravity theory with a Lagrangian $\mathcal{L}= \mathcal{L}(g_{\mu \nu}, R_{\alpha \beta}^{\mu \sigma})$,  the surface term is cast in the form 
\begin{equation}\label{surfaceterm}
    \Theta^{\alpha}(\delta g, \delta\Gamma)=2E^{\alpha \beta}_{\mu \sigma}\,\delta\Gamma^{\mu}_{\beta \lambda}g^{\lambda \sigma}+2\nabla^{\mu}E^{\alpha \beta}_{\mu \sigma}(g^{-1}\delta g)^{\sigma}_{\beta}\;,
\end{equation}
where the fourth-rank skew-symmetric tensor $E^{\alpha \beta}_{\mu \sigma}$ is defined as $E^{\alpha \beta}_{\mu \sigma}=\frac{\partial \mathcal{L}}{\partial R_{\alpha \beta}^{\mu \sigma}}$.  
Then the equations of motion can generally be written as
\begin{equation}
E_{\mu \nu }=E_{\mu }^{\ \lambda \rho \sigma }R_{\nu \lambda \rho \sigma }-
\frac{1}{2}\,g_{\mu \nu }\,\mathcal{L}-2\nabla ^{\lambda }\nabla ^{\rho
}E_{\mu \lambda \rho \nu }\,.
\end{equation}

For the particular case of QCG theory (\ref{QCGaction}), this tensor takes the form
\begin{equation}
E_{\mu \sigma }^{\alpha \beta}=\frac{1}{2\kappa }\,\delta _{\mu \sigma}^{\alpha \beta }+\frac{\alpha }{2}\,R_{[\mu }^{[\alpha }\delta _{\sigma]}^{\beta ]}+\beta R\,\delta _{\mu \sigma }^{\alpha \beta }+\frac{\gamma }{2}
\,\delta _{\mu \,\sigma \,\nu _{1}\,\nu _{2}}^{\alpha \,\beta \,\rho_{1}\rho _{2}}R_{\rho _{1}\rho _{2}}^{\nu _{1}\nu _{2}}\,.
\end{equation}

As a consequence, the surface term for the theory, defined in terms of the functional variation of the fields is
\begin{eqnarray}
    \Theta ^{\alpha }(\delta g, \delta\Gamma) &=&\left( \frac{1}{\kappa }\,\delta _{\mu \sigma }^{\alpha \beta }+\alpha \,R_{[\mu }^{[\alpha }\delta _{\sigma ]}^{\beta ]}+2\beta R\delta _{\mu \sigma }^{\alpha \beta }+\gamma\delta_{\mu\,\sigma\, \nu_{1}\,\nu_{2} }^{\alpha\,\beta\,\rho_{1} \rho_{2}}R^{\nu_{1}\nu_{2}}_{\rho_{1} \rho_{2}}\right) \,g^{\sigma \lambda }\delta \Gamma _{\lambda \beta }^{\mu }
    \notag \\
    &&+\nabla ^{\mu}\left( \alpha \,R^{[\alpha }_{[\mu }\delta ^{\beta ]}_{\sigma ]}+2\beta R\delta ^{\alpha \beta }_{\mu \sigma }\right) \left( g^{-1}\delta g\right) _{\beta }^{\sigma }\,.
\end{eqnarray}

Notice that the contributions coming from Einstein and Gauss-Bonnet terms in the bulk Lagrangian vanish identically under the covariant derivative in the second line.

Invariance under diffeomorphisms of the action (\ref{QCGaction}) implies that there is an associated Noether current $J^{\alpha}$, which is conserved on-shell $\nabla_{\alpha}J^{\alpha}=0$. For infinitesimal diffeomorphisms generated by a vector field $\xi=\xi^{\mu}\partial_{\mu}$, this current takes the general form
\begin{equation}
J^{\alpha}=\Theta^{\alpha}(\delta_{\xi} g, \delta_{\xi}\Gamma)+\mathcal{L}\,\xi^{\alpha}\,,
\end{equation}
\begin{equation}
J^{\alpha}=2\nabla_{\beta}\left(E^{\alpha \beta}_{\mu \sigma}\nabla^{\mu}\xi^{\sigma}+2\xi^{\mu}\nabla^{\sigma}E^{\alpha \beta}_{\mu \sigma} \right)\;.
\end{equation}

Since the current is conserved, the Poincaré lemma ensures the fact that $J^{\alpha}$ can be written locally as the divergence of a prepotential, $J^{\alpha}=\nabla_{\beta}Q^{\alpha \beta}$. However, it is the Noether-Wald procedure the one which allows to find explicitly such prepotential as 
\begin{equation}
Q^{\alpha \beta}=2\left(E^{\alpha \beta}_{\mu \sigma}\nabla^{\mu}\xi^{\sigma}+2\xi^{\mu}\nabla^{\sigma}E^{\alpha \beta}_{\mu \sigma} \right)=-Q^{\beta \alpha}\,.
\end{equation}

The Noether charge associated to the asymptotic Killing vector $\xi ^{i}$ on $\partial M $, with the topology $\mathbb{R}\times \Sigma$, is computed over the $(d-1)$-dimensional spatial surface $\Sigma $ with the surface element $d\Sigma _{\mu \nu }$, as
\begin{equation}
Q_{\textrm{NW}}[\xi ]=\int\limits_{\Sigma }d\Sigma _{\mu \nu }\,Q^{\mu \nu }(\xi )\,. \label{Q[xi]}
\end{equation}
In order to write the charge in its standard form, linear in the Killing vector, we use the unit normals on the co-dimension two hypersurface $\Sigma$, which are the timelike normal $u_{\mu }$ ($u_{\mu }u^{\mu }=-1$) and the spacelike one $n_{\mu }$ ($n_{\mu }n^{\mu }=1$). Denoting the metric on $\Sigma$ by $\sigma_{mn}$,  we can write the surface element as $d\Sigma _{\mu \nu }=\frac{1}{2}\,d^{d-1}x\sqrt{\sigma }\,n_{[\mu }u_{\nu ]}$. More details about conventions are given in Appendix \ref{Conventions}.

%%%%%%%%%%%%%%%%%%%%%%%%%%%%%%%%%%%%%%%%%%%%%%%%%%%%%%%%%%%%%%%%%

\section{Variational principle in odd dimensions}

Finite conserved charges in QCG were computed in even dimensions using the topological renormalization \cite{Giribet:2018hck,Giribet:2020aks}. In this case, a single topological term in the bulk plays the role of boundary counterterms. Therefore, the Wald
prescription can be applied for the full bulk gravity theory in a covariant fashion.

 Quite the opposite, in odd dimensions $d+1=2n+1$, there are no topological invariants of the Euler class. A possible way of circumventing this obstruction is to consider the addition of Kounterterms, as a consistent alternative to the renormalization of Einstein AdS gravity \cite{Olea:2006vd} and any EGB-AdS gravity theory \cite{Kofinas:2006hr}. In a similar manner, in what follows,
 a finite action principle for QCG in odd spacetime dimensions is obtained adding a boundary term of the type $B_{2n}(K, \mathcal{R})$.

The renormalized version of QCG theory considers the action in eq.\eqref{QCGaction}, plus a boundary term 

\begin{equation}
I^{\mathrm{ren}}_{\mathrm{QCG}}=I_{\mathrm{QCG}}+c_{2n}\int \limits_{\partial M} d^{2n}x\sqrt{-h}\,B_{2n},
\label{IrenQCG}
\end{equation}
which is a modification of the Kounterterms that appear in the Einstein gravity case in eq.\eqref{KountertermOdd}. Indeed, this modification amounts to the substitution in the AdS radius $\ell \to \ell_{\mathrm{eff}}$, that is,
%\begin{eqnarray}
%B_{2n} &=&2n\int\limits_{0}^{1}du\int\limits_{0}^{u}ds\,\delta
%_{i_{1}\cdots i_{2n}}^{j_{1}\cdots j_{2n}}K_{j_{1}}^{i_{1}}\delta
%_{j_{2}}^{i_{2}}\left(\frac{1}{2}\,\mathcal{R}_{j_{3}\,j_{4}}^{i_{3}\,i_{4}}-u^{2}K_{j_{3}}^{i_{3}}K_{j_{4}}^{i_{4}}+\frac{s^{2}}{\ell_{\mathrm{eff}}^2}\, \delta _{j_{3}}^{i_{3}}\delta _{j_{4}}^{i_{4}}\right) \times \cdots \notag \\
%&&\qquad \cdots \times \left( \frac{1}{2}\,\mathcal{R}_{j_{2n-1}%
%\,j_{2n}}^{i_{2n-1}\,i_{2n}}-u^{2}K_{j_{2n-1}}^{i_{2n-1}}K_{j_{2n}}^{i_{2n}}+\frac{s^{2}}{\ell_{\mathrm{eff}}^2}\,\delta _{j_{2n-1}}^{i_{2n-1}}\delta_{j_{2n}}^{i_{2n}}\right) \,. \label{Ktodd}
%\end{eqnarray}
\begin{equation}
 \mathcal{F}_{j\,l}^{i\,k}(u,s)=\frac{1}{2}\,\left(\mathcal{R}_{j\,l}^{i\,k}-u^{2}(K_{j}^{i}K_{l}^{k}-K_{l}^{i}K_{j}^{k})+\frac{s^{2}}{\ell_{\mathrm{eff}}^2}\,\delta _{jl}^{ik}\right)\,,    
\end{equation}
This correctly accounts for the modified asymptotic behavior induced by the quadratic curvature corrections in the action and field equations. The information on the bulk theory is also encoded in the constant $c_{2n}$, which has to be fixed so that a well posed action principle is attained. To this end, the variation of the renormalized QCG action  must vanish for boundary conditions compatible with asymptotically AdS spaces.

The on-shell variation of the bulk Lagrangian is always a total derivative, which can be written as a surface term by virtue of the Stokes' theorem. In Gaussian coordinates $x^{\mu }=(z,x^{i})$, the variation of the bulk action is
\begin{equation}
\delta I_{{\mathrm{QCG}}}=\int\limits_{\partial M}d^{2n}x\,\sqrt{-h}\,n_{\mu }\Theta ^{\mu }(\delta g,\delta \Gamma )=\int\limits_{\partial M}d^{2n}x\,N\sqrt{-h}\,\Theta ^{z}\,.
\label{Ibulk}
\end{equation}

As a consequence, the corresponding variation of the renormalized action for QCG is written as
\begin{equation} \label{deltaIren}
\delta I^{\rm{ren}}_{{\mathrm{QCG}}}=\int\limits_{\partial M}d^{2n}x\,\left(\sqrt{-h} N\Theta^z+c_{2n}\delta (\sqrt{-h}B_{2n})\right) \,.
\end{equation}
In order to uniform the expressions for $\Theta^z$ and $B_{2n}$ in the same coordinate (Gauss-normal) frame, we use the Christoffel symbols and the Gauss-Codazzi relations in Appendix \ref{foliation}. Thus, the expressions derived will be expressed in terms of the boundary tensorial quantities $h_{ij}, K_{ij}, \mathcal{R}_{jkl}^{i}$. The bulk variations $\delta g_{\mu \nu }$, $\delta \Gamma _{\nu \lambda }^{\mu }$ will be naturally mapped on the boundary to the variations $\delta h_{ij}$, $\delta K_{ij}$ ($\delta N=0$ on the boundary located at $z=const$). The variation $\delta \Gamma _{jk}^{i}(h)$ vanishes on the boundary by virtue of the discussion contained in Appendix \ref{SFF}. Simply put, variations of the Christoffel of the boundary metric always come in the form $\delta \theta_{jk}^{i}$, which vanishes identically.

In doing so, the variation of the bulk action takes the form \cite{Yoel}
\begin{eqnarray}
\delta I_{{\mathrm{QCG}}}\!\!\! &=&\!\!\!\!\!\int\limits_{\partial M}d^{2n}x\,\sqrt{-h}\, \left\{ \left( \left(h^{-1}\delta h\right)_{k}^{i}K_{j}^{k} +2\delta K_{j}^{i}\right)\left[ \frac{1}{\kappa }\delta_{i}^{j}+\left( \alpha R_{z}^{z}+2\beta R\right) \delta _{i}^{j}+\alpha R_{i}^{j}+\gamma \,\delta_{ipq}^{jmn}R_{mn}^{pq} \right] \right.   \notag \\
&&+\alpha N\nabla^{i}R_{[i}^{z}(h^{-1}\delta h)_{j]}^{j} +\left. N\nabla ^{z}\left[ \left(\alpha R_{z}^{z}+2\beta R\right) \delta _{j}^{i}+\alpha R_{j}^{i}\rule{0pt}{13pt} \right]  \left( h^{-1}\delta h\right) _{i}^{j} \rule{0pt}{15pt} \right\}.  \label{Ib}
\end{eqnarray}
%\begin{eqnarray}
%\delta I_{{\mathrm{QCG}}} &=&\int\limits_{\partial \mathcal{M}}d^{2n}x\,\sqrt{-h}\,\left\{ \left[ \left( h^{-1}\delta h\right) _{k}^{i_{1}}K_{j_{1}}^{k}+2\delta K_{j_{1}}^{i_{1}}\right] \left[ \frac{1}{\kappa }\,\delta _{i_{1}}^{j_{1}}+\left( \alpha \,R_{r}^{r}+2\beta R\right) \delta _{i_{1}}^{j_{1}}\right. \right.  \label{Ib}  \\
%&&+\left. \alpha \,R_{i_{1}}^{j_{1}}+\gamma \delta
%_{i_{1}i_{2}i_{3}}^{j_{1}j_{2}j_{3}} \left( \mathcal{R}_{j_{2}j_{3}}^{i_{2}i_{3}}-2K_{j_{2}}^{i_{2}}K_{j_{3}}^{i_{3}}\right) \right] +\alpha N\,R_{j}^{r}\,\delta \Gamma _{ik}^{[j}h^{i]k}+\alpha \,\nabla^{i}R_{[i}^{r}\left( h^{-1}\delta h\right) _{j]}^{j} \notag\\
%&&+\left. N\nabla ^{r}\left[ \left( \alpha \,R_{r}^{r}+2\beta R\right) \delta _{j}^{i}+\alpha \,R_{j}^{i}\right] \left( h^{-1}\delta h\right)_{i}^{j}+2\gamma N\,\delta _{i_{1}i_{2}i_{3}}^{j_{1}j_{2}j_{3}}\delta\Gamma_{kj_{1}}^{i_{1}}h^{i_{2}k}R_{j_{2}j_{3}}^{ri_{3}}\rule{0pt}{13pt}\right\}
%\,.   \notag
%\end{eqnarray}
%\mb{\begin{eqnarray}
%\delta  I_{bulk}&=&\int_{\partial\mathcal{M}}d^{2n}x\,\sqrt{-h}\left[ \rule{0pt}{13pt}(h^{-1}\delta h)^{i_{1}}_{m}K^{m}_{j_{1}}+2\delta K^{i_{1}}_{j_{1}}\right]\left[ \frac{1}{\kappa}\delta^{j_{1}}_{i_{1}} +\alpha\left(\delta^{j_{1}}_{i_{1}}R^{r}_{r}+R^{j_{1}}_{i_{1}}\right)+2\beta R\delta^{j_{1}}_{i_{1}}\right. \notag\\
%&&\left.\qquad+\gamma\delta^{j_{1} j_{2} j_{3}}_{i_{1}\,i_{2}\,i_{3}}R^{i_{2}\,i_{3}}_{j_{2}\,j_{3}}\rule{0pt}{17pt}\right].
%\end{eqnarray}}

In what follows, we consider asymptotically AdS spaces in QCG, where the induced metric has the pole of order two on the boundary. One can assume that this near-boundary behavior is a feature in common of both Einstein and non-Einstein branches in a higher-derivative gravity theory, even when new modes are switched on in the latter case. Therefore, solutions in both branches should be continuously connected to global AdS space, such that 
\begin{equation}\label{Rdelta}
R_{\alpha \beta }^{\mu \nu }=-\frac{1}{\ell _{\mathrm{eff}}^{2}}\,\delta _{\alpha \beta }^{\mu \nu }+\mathcal{O}(z^a)\,,
\end{equation}
where $a=2$ when $\alpha$ and $\beta$ vanish and $a=1$ in the generic higher-derivative case.

As a consequence, the extrinsic and intrinsic curvatures behave as
\begin{equation}\label{leading}
K_{j}^{i}=\frac{1}{\ell_{\mathrm{eff}}}\,\delta _{j}^{i}+\mathcal{O}(z^a)\,,\qquad \mathcal{R}_{kl}^{ij}=\mathcal{O}(z^a)\,,
\end{equation}
what justifies the fact that $\delta K^i_j=\mathcal{O}(z^a)$ is subleading with respect to $(h^{-1}\delta h)^i_j=\mathcal{O}(1)$ in the equation \eqref{Ib}\footnote{Should we have included the terms involving $\delta \Gamma_{kj}^{i}=\mathcal{O}{(1)}$, they would have proved to be subleading, due to the fact they always appeared multiplied by $NR_{jk}^{zi}=D_{[j}K_{k]}^{i} =\mathcal{O}{(z^a)}$, where the covariant derivative $D_i$ is defined with respect to the boundary metric.}. On the other hand, due to the asymptotic form of the spacetime curvature \eqref{Rdelta}, it is expected that covariant derivatives $\nabla _{z}$ or $\nabla _{i}$ acting on the Ricci tensor and Ricci scalar will give rise to the subleading terms in the second line of eq.\eqref{Ib}.  
%\begin{eqnarray}
%&&\frac{1}{\kappa }\,\delta _{i_{1}}^{j_{1}}+\left( \alpha
%\,R_{r}^{r}+2\beta R\right) \delta _{i_{1}}^{j_{1}}+\alpha
%\,R_{i_{1}}^{j_{1}}+\gamma \delta _{i_{1}i_{2}i_{3}}^{j_{1}j_{2}j_{3}}R_{j_{2}j_{3}}^{i_{2}i_{3}}%
 %  \notag \\
%&=&\left( \frac{1}{\kappa }-\frac{4n\alpha }{\ell _{\mathrm{eff}}^{2}}-\frac{4n\left( 2n+1\right) \beta }{\ell _{\mathrm{eff}}^{2}}-\frac{2\gamma \left(2n-1\right) \left( 2n-2\right) }{\ell _{\mathrm{eff}}^{2}}\right) \delta
%_{i_{1}}^{j_{1}}+\mathcal{O}_{2}\,,
%\end{eqnarray}

Therefore, in order to isolate the highest order divergent term $\mathcal{O}(z^{-2n})$ the leading order of the
bulk part \eqref{Ib} is
%\begin{eqnarray}
%\delta I_{{\mathrm{QCG}}} &=&\frac{1}{\ell _{\mathrm{eff}}}
%\int\limits_{\partial M}d^{2n}x\,\sqrt{-h}\,\left( h^{-1}\delta h\right) _{i}^{i}\,\times   \notag \\
%&&\times\left( \frac{1}{\kappa }-\frac{4n\alpha +4n\left( 2n+1\right) \beta+2\gamma \left( 2n-1\right) \left( 2n-2\right) }{\ell _{\mathrm{eff}}^{2}}+
%\mathcal{O}{(z^2)}\right) .
%\end{eqnarray}
\begin{eqnarray}
\delta I_{{\mathrm{QCG}}} &=&\int\limits_{\partial M}d^{2n}x\,\sqrt{-h}\left[
\rule{0pt}{15pt}\frac{1}{\ell _{\mathrm{eff}}}\,\left( h^{-1}\delta h\right)_{i}^{i}\right. \times   \notag \\
&&\times \left. \left( \frac{1}{\kappa }-\frac{4n\alpha +4n\left(
2n+1\right) \beta +2\gamma \left( 2n-1\right) \left( 2n-2\right) }{\ell _{\mathrm{eff}}^{2}}\right) +\mathcal{O}{(z^{a})}\right] .
\end{eqnarray}

At this level, we should emphasize that a proper holographic description of QCG is technically involved and lies beyond the scope of the present work. Instead, we will simply suggest that the addition of an appropriate boundary term $B_{2n}$ cures the theory from infrared divergences at the conformal boundary of AAdS spaces (from the next-to-leading order divergence to the holographic order). In the procedure below, the argument  is developed to the full extent based on the explicit asymptotic behavior of solutions to QCG.

On the other hand, the leading-order contribution from the boundary term is dictated by the relations  \eqref{leading}, such that

\begin{equation}
B_{2n} =\frac{2n}{\ell_{\mathrm{eff}}^{2n-1}}\delta
_{i_{1}\cdots i_{2n}}^{j_{1}\cdots j_{2n}}\delta_{j_{1}}^{i_{1}}\cdots\delta
_{j_{2n}}^{i_{2n}}\int\limits_{0}^{1}du\int\limits_{0}^{u}ds\,\left(s^{2}-u^{2}\,\right)^{n-1}+\mathcal{O}(z^{a})\,,
\end{equation}
or, equivalently,

\begin{equation}
\delta \left(c_{2n} \sqrt{-h}B_{2n}\right)=\sqrt{-h}\,\left( \left(
h^{-1}\delta h\right) _{i}^{i} \rule{0pt}{12pt}%
\frac{nc_{2n}}{\ell _{\mathrm{eff}}^{2n-1}}\,(-1)^{n-1}2^{2n-2}(n-1)!^{2}+\mathcal{O}{(z^{a})} \right).
\end{equation}

%\begin{equation}
%\delta B_{2n}=\frac{\sqrt{-h}}{\ell _{\mathrm{eff}}^{2n-1}}\,\left(
%h^{-1}\delta h\right) _{i}^{i}\left( \rule{0pt}{12pt}%
%n\,c_{2n}\,(-1)^{n-1}2^{2n-2}(n-1)!^{2}+\mathcal{O}{(z^2)}\right) ,
%\end{equation}
Here, we have performed explicitly the parametric integrations. This can be interpreted as the highest order divergence of the volume of the boundary hypersurface.

%So far we showed that $\Theta(\delta g, \delta\Gamma)=\Theta(\delta h, \delta K)$, and the same is valid for the boundary term, so that the on-shell variation has covariant form
%\begin{equation}
    %\delta I_{{\mathrm{QCG}}}=\int\limits_{\partial\mathcal{M}}d^{2n}x\,\mc{\cancel{N}}\sqrt{-h}\left(n_{\mu}\,\Theta^{\mu}(\delta h, \delta K)+\frac{c_{2n}}{\mc{\cancel{N}}\sqrt{-h}}\,\delta B_{2n}\right)\,,
%\end{equation}

A properly defined action principle would require all boundary terms in eq.\eqref{deltaIren} to vanish for boundary conditions compatible with the conformal boundary structure of AAdS spacetimes. As discussed above, a full proof based on the expansion of the fields in the vicinity of the boundary, is technically involved in higher-derivative gravity.  
This consideration implies, at the leading order, that the coupling of the boundary term appears to be fixed in terms of the parameters of the QCG theory, i.e.,
%\begin{eqnarray}
%\delta I^{\mathrm{ren}}_{{\mathrm{QCG}}} &=&\frac{1}{\ell _{\mathrm{eff}}}\left[ \frac{1}{%
%\kappa }-\frac{4n\alpha +4n\left( 2n+1\right) \beta }{\ell _{\mathrm{eff}%
%}^{2}}-\frac{2\gamma \left( 2n-1\right) \left( 2n-2\right) }{\ell _{\mathrm{%
%eff}}^{2}}\right.  \\
%&&+\left. \frac{nc_{2n}}{\ell _{\mathrm{eff}}^{2n-2}}%
%\,(-1)^{n-1}2^{2n-2}(n-1)!^{2}\right] \int\limits_{\partial
%M}d^{2n}x\,\sqrt{-h}\,\left(\left( h^{-1}\delta h\right) _{i}^{i}+\mathcal{O}{(z^{2})}\right).
%\notag
%\end{eqnarray}
%The expression vanishes  if the constant in front of the integral is identically zero, from where the coupling is found to be
\color{black}
\begin{equation} \label{acopleQCG}
c_{2n}=\frac{(-1)^{n}\ell _{\mathrm{eff}}^{2n-2}}{2^{2n-2}\kappa n(n-1)!^{2}}%
\left[ 1+\frac{4\kappa \Lambda _{\mathrm{eff}}}{(2n-1)}\left( \alpha +\left( 2n+1\right) \beta +\gamma \,\frac{\left( 2n-1\right) \left( 2n-2\right) }{2n}\right) \right] .
\end{equation}
%This coupling constant ensures a well-defined variational principle for asymptotic\mc{ally} AdS conditions.
%\vc{It also provides a regularized QCG action.}
%%%%%%%%%%%%%%%%%%%%%%%%%%%%%%%%%%%%%%%
\section{Noether theorem and energy}

Spacetime diffeomorphisms $\delta x^{\mu} =\xi^{\mu}(x)$ generate the changes of the dynamical fields in the gravitational theory. The invariance of the action \eqref{IrenQCG} under isometries defines a conservation law in terms of the Noether current, $\partial_{\mu}\left(\sqrt{-g}J^{\mu}\right)=0$. As reviewed in Section \ref{SecQCG}, this current considers a contribution from the bulk and, in addition, another which is coming from the diffeomorphic variation of $B_{2n}$, that is,
%\begin{eqnarray}
%\label{Noether}
%\delta(\sqrt{-h}B_{2n})
%&=&\sqrt{-h}\,\delta _{i_{1}\cdots i_{2n}}^{j_{1}\cdots j_{2n}}\left[
%\rule{0pt}{14pt}(h^{-1}\delta h)_{l}^{i_{1}}K_{j_{1}}^{l}+2\delta
%K_{j_{1}}^{i_{1}}\right] \delta _{j_{2}}^{i_{2}}\times \notag \\
%&& \times n\int\limits_{0}^{1}du\left( \frac{1}{2}\,R_{j_{3}\,j_{4}}^{i_{3}%
%\,i_{4}}+\frac{u^{2}}{\ell _{\mathrm{eff}}^{2}}\,\delta
%_{j_{3}}^{i_{3}}\delta _{j_{4}}^{i_{4}}\right) \cdots \left( \frac{1}{2}%
%\,R_{j_{2n-1}\,j_{2n}}^{i_{2n-1}\,i_{2n}}+\frac{u^{2}}{\ell _{\mathrm{eff}
%}^{2}}\,\delta _{j_{2n-1}}^{i_{2n-1}}\delta _{j_{2n}}^{i_{2n}}\right)
%\notag \\
%&&+\sqrt{-h}\,\delta _{i_{1}\cdots i_{2n}}^{j_{1}\cdots j_{2n}}\left[ \rule{0pt}{14pt}(h^{-1}\delta h)_{l}^{i_{1}}\left(
%K_{j_{1}}^{l}\delta_{j_{2}}^{i_{2}}-\delta_{j_{1}}^{l}K_{j_{2}}^{i_{2}}\right) +2\delta _{j_{1}}^{i_{1}}\,\delta
%K_{j_{2}}^{i_{2}}\right] \times \\
%&&\hspace{-1.5cm} \times n\int\limits_{0}^{1}du\,u\left( \frac{1}{2}\mathcal{R}_{j_{3}j_{4}}^{i_{3}
%i_{4}}-u^{2}K_{j_{3}}^{i_{3}}K_{j_{4}}^{i_{4}}+\frac{u^{2}}{\ell _{\mathrm{eff}}^{2}}\,\delta _{j_{3}}^{i_{3}}\delta _{j_{4}}^{i_{4}}\right) \cdots
%\left( \frac{1}{2}\mathcal{R}_{j_{2n-1}j_{2n}}^{i_{2n-1} i_{2n}}-u^{2}K_{j_{2n-1}}^{i_{2n-1}}K_{j_{2n}}^{i_{2n}}+\frac{u^{2}}{\ell_{\mathrm{eff}}^{2}}\,\delta _{j_{2n-1}}^{i_{2n-1}}\delta
%_{j_{2n}}^{i_{2n}}\right) .  %\notag
%\end{eqnarray}
\begin{eqnarray}
\label{Noether}
\delta_{\xi}(\sqrt{-h}B_{2n})
&=&\sqrt{-h}\,n\,\delta _{i_{1}\cdots i_{2n}}^{j_{1}\cdots j_{2n}}\left[
\rule{0pt}{14pt}(h^{-1}\delta_{\xi} h)_{l}^{i_{1}}K_{j_{1}}^{l}+2\delta_{\xi}
K_{j_{1}}^{i_{1}}\right] \delta _{j_{2}}^{i_{2}}
 \int\limits_{0}^{1}du\,\mathcal{F}_{j_{3}\,j_{4}}^{i_{3}\,i_{4}}(1,u) \cdots 
\,\mathcal{F}_{j_{2n-1}\,j_{2n}}^{i_{2n-1}\,i_{2n}}(1,u)
\notag \\
&&+\sqrt{-h}\,n\, \delta _{i_{1}\cdots i_{2n}}^{j_{1}\cdots j_{2n}}\left[ \rule{0pt}{14pt}(h^{-1}\delta_{\xi} h)_{l}^{i_{1}}\left(
K_{j_{1}}^{l}\delta_{j_{2}}^{i_{2}}-\delta_{j_{1}}^{l}K_{j_{2}}^{i_{2}}\right) +2\delta _{j_{1}}^{i_{1}}\,\delta_{\xi}
K_{j_{2}}^{i_{2}}\right]\times \notag \\
&&\times\int\limits_{0}^{1}du\,u\,\mathcal{F}_{j_{3}j_{4}}^{i_{3}
i_{4}}(u,u) \cdots
\mathcal{F}_{j_{2n-1}j_{2n}}^{i_{2n-1} i_{2n}}(u,u)\,,
\end{eqnarray}
which is a kind of involved expression.

%%J^{\mu}=\Theta^{\mu}(\delta h, \delta K)+\mathcal{L}\xi^{\mu}+c_{2n}\delta_{\xi}B_{2n}\,, \mc{\qquad \text{(erase)}}
%\end{equation}
%\begin{equation}\label{Noethercurrentmodified}
     %J^{\mu}=\Theta^{\mu}+\mathcal{L}\xi^{\mu}+\frac{c}{\sqrt{-h}}\,n^{\mu}\partial_{i}\left(\xi^{i}B_{2n}\right)+c_{2n}n^{\mu}\pounds_{\xi}B_{2n}\,, \mc{\qquad \text{(erase)}}
%\end{equation}
Fortunately, as the diffeomorphic variations are written in terms of Lie derivatives along the vector $\xi^{\mu}$, the extra term in the Noether current can be written down as
\begin{equation}  \label{J}
J^{\mu }=\Theta ^{\mu }(\delta h,\delta K)+\mathcal{L}\xi ^{\mu }+\frac{c_{2n}\,n^{\mu }}{\sqrt{-h}}\,\partial _{i}\left(\sqrt{-h}\,\xi ^{i}
B_{2n}\right) \,.
\end{equation}

As covariance is manifestly broken by the inclusion of a boundary term, the formulas for the conserved quantities below are expressed in Gaussian coordinates \eqref{NormalCoordinates}. 

The conservation of the Noether current  defines a constant of motion in the radial direction
\begin{equation} \label{Q(J)}
    Q[\xi]=\int\limits_{\partial M} d^dx\,\sqrt{-h}\,n_{\mu}J^{\mu}\,.  
\end{equation}

While the fact the current is conserved implies that it may be written as a total derivative in a coordinate patch, here this is true globally. Thus, the current projected to the boundary can be written as a divergence of a prepotential $Q^{\mu \nu}$, i.e.,  $n_{\mu}J^{\mu}=\nabla_{\nu}(n_{\mu} Q^{\mu\nu})$. Furthermore, since the prepotential is linear in the asymptotic Killing vector,
\begin{eqnarray} 
    Q[\xi]&=&\int\limits_{\Sigma} d^{d-1}x\,\sqrt{\sigma}\,n_{\mu}u_{\nu}Q^{\mu\nu}\,,\\
    &=&\int\limits_{\Sigma} d^{d-1}x\,\sqrt{\sigma}u_i\,(q^{i}_{j}+q^{i}_{(0)j})\, \xi^j\,,
\end{eqnarray}
in terms of two charge density tensors,
%$ The prepotential in odd dimensions according to (\ref{Prepotencialmod}), takes the form
%$$q^{\mu \nu}(\xi)\;\longrightarrow\;q^{\mu \nu}(\xi)+\xi^{[\mu}B^{\nu]}\,,$$
%\begin{equation}
%       =\left(q^{\mu \nu}(\xi)+q^{\mu \nu}_{0}(\xi)\right)+\xi^{[\mu}B^{\nu]}\,,
%\end{equation}
%\begin{eqnarray}
%\mc{Q}^{\mu \nu }(\xi ) &\to &\mc{Q}^{\mu \nu }(\xi )+\xi ^{[\mu }B^{\nu ]}  \notag \\
%&=&\mc{Q}^{\mu \nu }(\xi )+\mc{Q}_{0}^{\mu \nu }(\xi )+\xi ^{[\mu }B^{\nu ]}\,,
%\end{eqnarray}
%where $B^{\mu}=B^{r}$ is the Kounterterm (\ref{Ktodd}), and the prepotential $\mc{Q}^{\mu \nu}$ is separated in two terms (plus the boundary correction term)....
%\begin{equation}
%Q[\xi ]=\int\limits_{\Sigma}d^{d-1}x\,\sqrt{\sigma }\,u_{i}\,(q^{i}_{j}+q^{i}_{(0)j})\,\xi
%^{j}\,,  \label{chargeproyected}
%\end{equation}
where $u_{\mu }=(0,u_{i})$. %One arrives to the same expression directly from \eqref{Q[xi]} by writing the surface element as $d\Sigma _{\mu \nu }=\frac{1}{2}\,d^{d-1}x\sqrt{\sigma }\,u_{[\mu }n_{\nu ]}$.

The first one has the form, 
\begin{eqnarray}
\label{chargeoddQCG}
q_{i}^{j} &=&\frac{1}{2^{n-2}}\,\delta _{ki_{1}\cdots
i_{2n-1}}^{jj_{1}\cdots j_{2n-1}}\,K_{i}^{k}\delta _{j_{1}}^{i_{1}}\left[
nc_{2n}\int\limits_{0}^{1}du\left( R_{j_{2}j_{3}}^{i_{2}i_{3}}+\frac{u^{2}}{\ell _{\mathrm{eff}}^{2}}\,\delta _{j_{2}j_{3}}^{i_{2}i_{3}}\right) \cdots
\left( R_{j_{2n-2}j_{2n-1}}^{i_{2n-2}i_{2n-1}}+\frac{u^{2}}{\ell _{\mathrm{eff}}^{2}}\,\delta _{j_{2n-2}j_{2n-1}}^{i_{2n-2}i_{2n-1}}\right) \right.
\notag \\
&&-\left. \frac{1}{(2n-1)!}\left( \frac{1}{\kappa }\,\delta
_{j_{2}j_{3}}^{i_{2}i_{3}}+\alpha \,R_{z}^{z}\delta
_{j_{2}j_{3}}^{i_{2}i_{3}}+2\beta \,R\delta
_{j_{2}j_{3}}^{i_{2}i_{3}}+2(2n-1)(2n-2)\gamma
R\,_{j_{2}j_{3}}^{i_{2}i_{3}}\right) \delta _{j_{4}j_{5}}^{i_{4}i_{5}}\cdots
\delta _{j_{2n-2}j_{2n-1}}^{i_{2n-2}i_{2n-1}}\rule{0pt}{20pt}\right]   \notag
\\
&&-2N\left[ \rule{0pt}{15pt}\alpha \left( \nabla ^{z}R_{i}^{j}-\nabla
^{j}R_{i}^{z}+\nabla ^{z}R_{z}^{z}\delta _{i}^{j}+\nabla ^{k}R_{k}^{z}\delta
_{i}^{j}\right) +2\beta \,\nabla ^{z}R\,\delta _{i}^{j}\right] -2\alpha
\,K_{k}^{j}R_{i}^{k}  \notag \\
&&-\ \, N\left[ \rule{0pt}{15pt}\alpha \left( \nabla ^{k}R_{k}^{z}\delta_{i}^{j}-\nabla ^{j}R_{i}^{z}\right) +4\gamma \left( \nabla
^{j}R_{i}^{z}-\delta _{i}^{j}\nabla ^{k}R_{k}^{z}+\nabla
^{k}R_{ik}^{jz}\right) \right] \,,
\end{eqnarray}
where we have used Gauss-Codazzi relations to write down
\begin{equation}
\mathcal{F}_{j\,l}^{i\,k}(1,u)=\frac{1}{2}\left(R_{j\,l}^{i\,k}+u^2\,\delta_{j\,l}^{i\,k}\right)\,.
\end{equation}
The charge tensor provides a notion of the energy and, eventually, other conserved quantities of the theory,
\begin{equation}\label{cargacod2}
    q[\xi]=\int\limits_{\Sigma}d^{d-1}x\,\sqrt{\sigma}\, u_{i}\,q^{i}_{j}\,\xi^{j}\,,
\end{equation}
In addition, we obtain an extra contribution coming from the surface term in the action  
\begin{eqnarray}
q_{(0)j}^{i} &=&nc_{2n}\,\delta _{k j_{1}\cdots j_{2n-1}}^{i i_{1}\cdots i_{2n-1}}\int\limits_{0}^{1}du\,u\left(
K_{j}^{k}\delta _{i_{1}}^{j_{1}}+K_{i_{1}}^{k}\delta _{j}^{j_{1}}\right) \left( \frac{1}{2}\,\mathcal{R}_{i_{2}i_{3}}^{j_{2}j_{3}}-u^{2}K_{i_{2}}^{j_{2}}K_{i_{3}}^{j_{3}}+\frac{u^{2}}{\ell _{\textrm{eff}}^{2}}\,\delta _{i_{2}}^{j_{2}}\delta_{i_{3}}^{j_{3}}\right) \times   \notag \\
&&\qquad \cdots \times \left(  \frac{1}{2}\,\mathcal{R}_{i_{2n-2}\,i_{2n-1}}^{j_{2n-2}\,j_{2n-1}}-u^{2}K_{i_{2n-2}}^{j_{2n-2}}K_{i_{2n-1}}^{j_{2n-1}}+\frac{u^{2}}{\ell _{\textrm{eff}}^{2}}\,\delta _{i_{2n-2}}^{j_{2n-2}}\delta_{i_{2n-1}}^{j_{2n-1}}\right) \,.  \label{chargevacuumQCG} 
\end{eqnarray}

Because the formula \eqref{chargeoddQCG} is such that it identically vanishes for global AdS space, the latter equation gives rise to the vacuum energy of AAdS gravity
\begin{equation}\label{cargacod2vacuum}
    q_{(0)}[\xi]=\int\limits_{\Sigma} d^{d-1}x\,\sqrt{\sigma}\,u_{i}\,q_{(0)j}^{i}\,\xi^{j}\,.
\end{equation}

Some of the properties of these quantities are made more evident when evaluated on an explicit AAdS black hole solution.

%%%%%%%%%%%%%%%%%%%%%%%%%%%%%%%%%%%%%%%%%%%%%%%%%%%%%%%%%%
\section{Energy of Topological Static Black Holes} \label{EnergyTBH}

Black hole solutions are an essential feature of gravitational theories, first observed in the context of Einstein gravity.  Therefore, it is expected that higher-curvature theories also posses geometries with a horizon as a solution to the corresponding field equations, most of them yet to be found \cite{Kehagias:2015ata,Stelle:2017bdu}.

 In order to check the charge formulas derived in the previous section, we consider static black hole geometries with topological horizon, which appear naturally in AdS gravity \cite{Birmingham:1998nr}. The condition of a static ansatz implies a line element $ds^2=g_{\mu \nu}(x)dx^{\mu}dx^{\nu}$ which, in the coordinate frame  $x^{\mu} = (t, r, \varphi^{n})$, takes the form
\begin{equation}\label{ansatzTBH}
    ds^{2}=-f^{2}(r)dt^{2}+f^{-2}(r)dr^{2}+r^{2}\gamma_{m n}(\varphi)\,d\varphi^{m}d\varphi^{n}\,,\qquad\varphi^{m} \in \Sigma^{d-1}\,.
\end{equation}
Here, $\gamma_{m n}(\varphi)$ is the metric of a $(d-1)$-dimensional Riemann space $\Sigma^{d-1}$ with constant curvature $k$,
\begin{equation}\label{constanttrasnversalcurvature}
    \mathcal{R}^{mn}_{pq}(\gamma)=k\,\delta^{mn}_{pq}
\end{equation}
such that $k= +1, 0 $ or $-1$ locally describes a spherical, flat or hyperbolic transversal section, respectively. The spacetime in between the horizon and infinity is foliated by a radial coordinate adapted to the topology of the transversal section. In particular, the horizon $r = r_{0}$ (defined by the largest root of the equation $f(r_{0}) = 0$) has the same topology.

For a topological black hole (\ref{ansatzTBH}), the nonzero  components of the extrinsic curvature take the simple form
\begin{equation}
 K^{t}_{t}=-f'\;,\;\;\;\;\;\;\;K^{n}_{m}=-\frac{f}{r}\,\delta^{n}_{m}\;,\;\;\;\;\;\;\;%\mb{K=-f'-\frac{(d-1)f}{r}}\,,
\end{equation}
while the trace $K=h^{i j}K_{i j}$ is
\begin{eqnarray}
 K=-f'-\frac{(d-1)f}{r}\,.
\end{eqnarray}
Also, the only non-vanishing components of the intrinsic curvature of the boundary metric are related to eq.(\ref{constanttrasnversalcurvature}) by a conformal rescaling, i.e.,
\begin{equation}
    \mathcal{R}^{i j}_{k l}(h)=\frac{k}{r^{2}}\,\delta^{i j}_{k l}\,.
\end{equation}

In order to evaluate the energy for the static black hole \eqref{ansatzTBH}, it may be convenient to have the components of the spacetime Riemann tensor
\begin{equation}
    R^{t n}_{t m}=R^{r n}_{r m}=-\frac{1}{2r}\,(f^{2})'\,\delta^{n}_{m}\,,
    \qquad R^{t r}_{t r}=-\frac{1}{2}\,(f^{2})'',\qquad 
    R^{m n}_{k l}=\frac{(k-f^{2})}{r^{2}}\,\delta^{mn}_{kl}\,,
\end{equation}
together with the Ricci tensor 
\begin{equation}
    \begin{aligned}
    R^{t}_{t}=R^{r}_{r}=-\frac{1}{2r}\left(r(f^{2})''+(d-1)(f^{2})' \right)\,,\\
    R^{n}_{m}=-\frac{1}{r^{2}}\left(r(f^{2})'-(d-2)(k-f^{2})\right)\delta^{n}_{m}\,,
    \end{aligned}
\end{equation}
and Ricci scalar
\begin{equation}
    R=-\frac{1}{r^{2}}\left(r^{2}(f^{2})''+2r(d-1)(f^{2})'-(d-1)(d-2)(k-f^{2})\right)\,.
\end{equation}

In four-dimensional gravity, the curvature-squared couplings do not modify the value of the effective AdS radius respect to the one defined by the bare cosmological constant $\Lambda_{0}<0$. Thus, it is expected that the solutions to QCG will behave asymptotically  \cite{Svarc:2018coe, Klemm:1998kf, Ogushi:2001gp} 
\begin{equation}
    f^{2}(r)= k-\frac{r_{0}}{r}+\frac{r^{2}}{\ell^{2}}\,.
\end{equation}
On the other hand,  Einstein-Gauss-Bonnet gravity in an arbitrary dimension ($\alpha=0=\beta$) has an exact Boulware-Deser solution \cite{Boulware:1985wk,Deser:2002jk}, where the near-boundary expansion of the function in the metric is
\begin{equation}
   f_{\pm}^{2}(r)\approx k-\left(\frac{r_{0}}{r}\right)^{d-2}+\frac{r^{2}}{\ell_{\pm}^{2}}+\cdots\,,
\end{equation}
where $\ell^{2}_{\pm}$ correspond to two effective AdS radii. As a consequence, even though there is no generic static black hole solution in QCG, one can assume an asymptotic behavior of the form \cite{Deser:2002jk}
\begin{equation}\label{SchQCG}
    f^{2}(r)\approx k-\left(\frac{r_{0}}{r}\right)^{d-2}+\frac{r^{2}}{\ell^{2}_{\mathrm{eff}}}+\cdots\;,
\end{equation}
whenever the AdS vacuum of the theory is not degenerate.

Therefore, we may evaluate the conserved charge associated to asymptotic Killing vectors in odd-dimensional QCG in eqs.(\ref{cargacod2}) and  (\ref{cargacod2vacuum}). With these formulae in mind, we can express the energy of an asymptotically AdS black hole  in QCG (\ref{SchQCG}), using the timelike normal vector $u_{i}=(f,0)$  and the metric on $\Sigma $ given by $\sigma _{mn}=r^{2}\gamma_{mn}(\varphi)$, in the form 
\begin{equation}\label{qodd1}
    E=q[\partial_{t}]=\left[1+\frac{4\kappa\Lambda_{\mathrm{eff}}}{(2n-1)}\left(\alpha+\beta(2n+1)+\gamma\frac{(2n-2)(2n-3)}{2n}\right)\right]\frac{\mathrm{Vol}(\Sigma^{2n-1})}{\mathrm{Vol}(S^{2n-1})}\frac{(2n-1)r_{0}^{2n-2}}{4G}\,,
\end{equation}
where $\mathrm{Vol}(\Sigma^{2n-1} )$ is the volume of the $(2n-1)$-dimensional transversal section, defined by the integral
    \begin{equation}
\mathrm{Vol}(\Sigma^{2n-1} )=\int d^{2n-1}\varphi\,\sqrt{\gamma }\,.
\end{equation}
The expression (\ref{qodd1}) can be equivalently written as 
\begin{equation}\label{qodd2}
    E=\left[-1+\frac{8\kappa\Lambda_{\mathrm{eff}}}{(2n-1)^{2}}\left(\alpha+(2n+1)\beta\right)+\frac{2\Lambda_{0}}{\Lambda_{\mathrm{eff}}} \right]\frac{\mathrm{Vol}(\Sigma^{2n-1})}{\mathrm{Vol}(S^{2n-1})}\frac{(2n-1)r_{0}^{2n-2}}{4G}\,,
\end{equation}
by substituting the factor $\gamma$ in terms of the other couplings of the theory with the help of eq.(\ref{comparacionLambda}).

In a similar fashion, the charge formula  (\ref{cargacod2vacuum}) gives rise to the vacuum energy
\begin{equation}
E_{\mathrm{vac}}=q_{(0)}[\partial_{t}]=k^{n}\left( 2n-1\right) !\,c_{2n}\,\mathrm{Vol}(\Sigma^{2n-1})\,,
\end{equation}
what is equivalent to evaluating the metric function $f^{2}(r)$ for pure AdS space.
Plugging in the constant (\ref{acopleQCG}), the vacuum energy of topological
black hole is \footnote{Due to the convention used here, the coupling in front of the EH term is $\kappa=2\textrm{Vol}(S^{2n-1})G$. Therefore, the vacuum the energy does not match the usual result for, e.g.,  Einstein-AdS gravity in five bulk dimensions ($\alpha=\beta=\gamma=0$), $E_{\mathrm{vac}}=\frac{3\pi\ell^{2}}{32 G}$. In our case, this expression produces $E_{\mathrm{vac}}=\frac{3\ell^{2}}{8 G}$.}
\begin{equation}
E_{\mathrm{vac}}=(-k)^{n}\frac{2(2n-1)!!^{2}}{(2n)!}\frac{\mathrm{Vol}(\Sigma^{2n-1})}{\kappa}\ell^{2n-2}_{\mathrm{eff}}\left[1-\frac{4n\kappa}{\ell^{2}_{\mathrm{eff}}}\left(\alpha+(2n+1)\beta+\gamma\frac{(2n-1)(2n-2)}{2n} \right) \right]\,.
\end{equation}

%%%%%%%%%%%%%%%%%%%%%%%%%%%%%%%%%%%%%%%%%%%%%%%%%%%%%%%%%5
\section{Conclusions}

In the present work, a formula for the conserved quantities in Quadratic Curvature Gravity in odd dimensions for the AAdS sector was obtained. The evaluation of the Noether-Wald charge for static black hole geometries, which are asymptotic solutions of QCG, gives rise to divergent results when taking the surface integral to radial infinity. It is therefore quite relevant to have a proper discussion of boundary terms and boundary conditions, which render the action principle finite in AAdS spacetimes. This is achieved by the addition of Kounterterms as boundary terms \cite{Olea:2006vd}, which are consistent with the holographic description of the gravity theory \cite{Miskovic:2009bm, Anastasiou:2020zwc}.

The energy of AAdS black holes, in particular, matches the results in ref.\cite{Deser:2002jk}, derived from the linearization of the field equations. In contrast, the obtention of the vacuum energy would require the use a background-independent method of computation. In that respect, a clear advantage of the procedure developed here is the fact that an expression for $E_{\rm{vac}}$ is given for any odd bulk dimension. Furthermore, this result reduces to the corresponding one in Einstein-Gauss-Bonnet AdS gravity \cite{Kofinas:2006hr}. We may
emphasize  that the conserved charge in eq.\eqref{chargeoddQCG} is free from the drawbacks found in perturbative/linearized frameworks, as they are in general not applicable at the points in the parameter space where the theory features a degenerate vacuum. 

The basic assumption is that the asymptotic behavior of the AAdS solutions of QCG is the same as the one of global AdS (vacuum state). Therefore, the modes switched on in generic spacetimes are such they induce perturbations in the spacetime curvature which are renormalizable by the addition of counterterms of the extrinsic type. It would be interesting, therefore, to study conserved quantities for non-Einstein solutions in higher-order gravity theories \cite{Alishahiha:2011yb,Cai:2009ac}.

Another relevant prospect is to work out a holographic description of QCG AdS, what includes reproducing the corresponding conformal anomalies \cite{Ghodsi:2019xrx}. We hope to report on this issue elsewhere.

\section*{Acknowledgments}
This work was funded in part by the grants FONDECYT N$^{\circ }$1190533 {\it Black holes and asymptotic symmetries}, VRIEA-PUCV N$^{\circ }$123.764, and ANID-SCIA-ANILLO ACT210100 {\it Holography and its applications to High Energy Physics, Quantum Gravity and Condensed Matter Systems}. YPC is supported by the ANID National M.Sc.~Scholarship N$^{\circ }$ 22210771.

\appendix

%%%%%%%%%%%%%%%%%%%%%%%%%%%%%%%%%%%%%%%%%%%%%%%%%%%%
\section{Conventions and identities}
\label{Conventions}

\subparagraph{{\it i}) Kronecker delta of order $p$.}

In many derivations throughout this work, we have used the totally antisymmetric Kronecker delta of order $p$, 
\begin{equation}
\delta^{\mu_{1} \mu_{2}\cdots\mu_{p}}_{\nu_{1} \nu_{2}\cdots\nu_{p}}= \det[\delta^{\mu_{1}}_{\nu_{1}}\cdots\delta^{\mu_{p}}_{\nu_{p}}]\,,
\end{equation}
which satisfies the general property of contraction of $k$ indices,
\begin{equation}
\delta^{\mu_{1}\cdots\mu_{k}\cdots\mu_{p} }_{\nu_{1}\cdots\nu_{k}\cdots\nu_{p} }\delta^{\nu_{1}}_{\mu_{1}}\cdots\delta^{\nu_{k}}_{\mu_{k}}=\frac{\left(N-p+k\right)!}{\left(N-p\right)!}\, \delta^{\mu_{k+1}\cdots\mu_{p} }_{\nu_{k+1}\cdots\nu_{p}}\,, 
\end{equation}
 where  $k\leq p$ and $N$ is the range of the indices.

This antisymmetric Kronecker tensor allows us to write down terms in a more compact form. For example,  the  Lanczos tensor  (\ref{Lanczos}) can  be rewritten as 
\begin{equation}\label{LanczosDelta}
    H^{\mu}_{\nu}=-\frac{1}{8}\,\delta _{\nu \nu _{1}\cdots \nu _{4}}^{\mu \mu _{1}\cdots \mu _{4}}R_{\mu _{1}\mu _{2}}^{\nu _{1}\nu _{2}}R_{\mu _{3}\mu _{4}}^{\nu _{3}\nu _{4}}\,.
\end{equation}

\subparagraph{{\it ii}) Useful integrals.} 
Starting from the known integral
\begin{equation}\label{integral1}
\int\limits_{0}^{1}du\,\left( u^{2}-1\right) ^{n-1}=\frac{(-1)^{n-1}2^{2n-2}(n-1)!^{2}}{(2n-1)!}\,,
\end{equation}
the double integral appearing in $\delta B_{2n}$ can be solved by the change of the variable $s\to us$, with the result
\begin{equation}\label{integral2}
\int\limits_{0}^{1}du\int\limits_{0}^{u}ds\,\left[ \rule{0pt}{10pt}s^{2}-(2n-1) u^{2}\right] \left(s^{2}-u^{2}\right) ^{n-2}=\frac{(-1)^{n-1}2^{2n-1}n(n-1)!^{2}}{(2n)!}\,.
\end{equation}
Other integrals used in the text are
\begin{equation}
\int\limits_{0}^{1}du\int\limits_{0}^{t}ds\left(
s^{2}-u^{2}\right)
^{n-1}=\frac{1}{2n}\frac{(-1)^{n-1}2^{2n-2}(n-1)!^{2}}{(2n-1)! 
}\,,  \label{s2-t2}
\end{equation}
and
\begin{equation}
\int\limits_{0}^{1}du\,\left[ a+(2n-1)\,bu^{2}\right] \left( a+u^{2}b\right)
^{n-2}=\left( a+b\right) ^{n-1}\,,\qquad n\geq 2\,.  \label{integral(2n-1)}
\end{equation}

\subparagraph{{\it iii}) Co-dimension two surface element.}

The surface element of ($d-1$)-dimensional manifold $\Sigma $  of co-dimension two, with the induced metric $\sigma _{mn}$ and the unit normals $u_{\mu }$ (time-like) and $n_{\nu }$ (space-like) in $M$, is given by
\begin{equation}
d\Sigma _{\mu \nu }=\frac{1}{2}\,d^{d-1}\sqrt{\sigma }\,n_{[\mu }u_{\nu ]}\,.
\label{dSigma}
\end{equation}
On the other hand, the pre-potential $Q^{\mu \nu }=-Q^{\nu \mu }$ is obtained from the Noether current,
\begin{equation}
n_{\mu }J^{\mu }=\nabla _{\nu }\left( n_{\mu }Q^{\mu \nu }\right) \,. \label{preQ}
\end{equation}
Thus, the charge in terms of the prepotential acquires the form
\begin{eqnarray}
Q[\xi ] &=&\int\limits_{\partial M}d^{d}x\sqrt{-h}\,\nabla _{\nu }\left(
n_{\mu }Q^{\mu \nu }\right)   \notag \\
&=& \int\limits_{\partial M}d^{d-1}x\sqrt{\sigma }\,n_{\mu }u_{\nu }Q^{\mu \nu }=\int\limits_{\Sigma }d\Sigma _{\mu \nu }Q^{\mu \nu
}\,.
\end{eqnarray}
Furthermore, the explicit $\xi $-dependence is introduced linearly by the charge density tensor,
\begin{equation}
n_{\nu }Q^{\nu \mu }=(q_{\lambda }^{\mu }+q_{(0)\lambda }^{\mu })\,\xi ^{\lambda }+\textrm{cov.~divergence},
\end{equation}
leading to the charge in the form used in Section \ref{Noether}. 
%%%%%%%%%%%%%%%%%%%%%%%%%%%%%%%%%%%%%%%%%%%%%%%%%%%%
\section{\texorpdfstring{$d+1$}{d+1} foliation of the spacetime}
\label{foliation}

The foliation \eqref{NormalCoordinates} for the spacetime, in terms of Gaussian coordinates, implies the following form for the different components of the connection $\Gamma^{\alpha}_{\mu \nu}$ 
\begin{equation} \label{ChristoffelK}
    \begin{split}
        &\Gamma ^{z}_{ij}=\frac{1}{N}\,K_{ij}\,, \qquad \, \Gamma ^{z}_{zz} = \frac{1}{N}\,\partial_{z}N\,,\\
        &\Gamma ^{i}_{z j} =-NK^{i}_{j}\,, \qquad\Gamma^{i}_{j k}=\Gamma^{i}_{j k}(h)\,.\\
    \end{split}
\end{equation}
 On the r.h.s.~of the expressions, boundary indices are lowered and raised  with the induced metric. $\Gamma^{i}_{j k}(h)$ is Christoffel symbol of the hypersurface $\partial M$.   

The choice of normal coordinates also leads to the Gauss-Codazzi relations for the spacetime curvature
\begin{eqnarray}
R_{kl}^{ij} &=&\mathcal{R}_{kl}^{ij}(h)-K_{k}^{i}K_{l}^{j}+K_{l}^{i}K_{k}^{j}\,, \notag \\
R_{jk}^{zi} &=& \frac{1}{N}\left( D_{j}K_{k}^{i} -D_{k}K_{j}^{i}\right) \,,
\label{Gauss-Codazzi} \\
R_{zj}^{zi} &=&\frac{1}{N}\, \partial _{z} K_{j}^{i}-K_{l}^{i}K_{j}^{l}\,,
\notag
\end{eqnarray}
where $D_{i}A^{l}_{m}=\partial_{i}A^{l}_{m}+\Gamma^{l}_{i k}(h) A^{k}_{m}-\Gamma^{k}_{i m}(h) A^{l}_{k}$ is the covariant derivative associated with induced geometry. 

%%%%%%%%%%%%%%%%%%%%%%%%%%%%%%%%%%%

\section{Second Fundamental Form}
\label{SFF}

The construction of boundary terms which depend\mc{s} on the Christoffel symbol requires precise matching conditions for a cobordant metric, in order to restore covariance at the boundary \cite{Eguchi:1980jx}. Indeed, in the Gaussian frame (\ref{NormalCoordinates}), it is evident that part of $\Gamma^{\alpha}_{\mu \nu}$ is a connection for the boundary metric.

In order to extract the tensorial content of the Christoffel symbol, one may define the second fundamental form as the difference between two connections, i.e., $\theta^{\mu}_{\alpha \beta}=\Gamma^{\mu}_{\alpha \beta}-\bar{\Gamma}^{\mu}_{\alpha \beta}$. The key point is to consider a second connection which comes from a product metric 
\begin{equation}\label{productmetric}
   d\bar{s}^2 = N^{2}(z)dz^{2}+\bar{h}_{i j}(x)dx^{i}dx^{j}\,,
\end{equation}
such that it satisfies the matching condition $\bar{h}_{i j}(x)=h_{i j}(z=z_{B}, x)$, where $z_B=const$ corresponds to the boundary.  The nonzero components of the Christoffel symbol are
\begin{equation}
    \label{Christoffel2}
        \bar{\Gamma} ^z_{zz}=\frac{1}{N}\,\partial_{z}N(z)\,,\;\;\;\;\;\;\;\bar{\Gamma}^{i}_{j k}=\Gamma^{i}_{j k}(\bar{h})\,,
\end{equation}
such that, at the boundary, 
%\mb{\begin{equation}
    %\label{tensorTheta}
       % \theta^{r}_{i j}=\frac{1}{N}K_{i j}\;,\;\;\;\;\;\;\;\;\theta^{i}_{r j}=-NK^{i}_{j}\,.
%\end{equation}}
%\mb{In turn, we have to the nonzero component of the tensor
%$\theta^{\mu\;\alpha}_{\;\;\,\beta}=g^{\alpha \lambda}\theta^{\mu}_{\;\lambda \beta}$}
\begin{eqnarray}
&&\theta^{zi}_{\;\;\,j}=g^{i k}\theta^{z}_{\;k j}=\frac{1}{N}\,K^{i}_{j}\,,\notag  \\
&&\theta^{iz}_{\;\;\,j}=g^{zz}\theta^{i}_{\;z j}=-\frac{1}{N}\,K^{i}_{j}\,, \\
&&\theta^{ij}_{\;\;\,z}=g^{j k}\theta^{i}_{\;k z}=-NK^{i j}\,,\notag\\
&&\theta^{ij}_{\;\;\,k}=0\, .\notag   
\end{eqnarray}
With all these ingredients, one can write down  Gauss-Codazzi relations in terms of the tensor $\theta^{\mu\alpha}_{\;\;\,\beta}$ on the boundary as
\begin{equation}\label{Gauss-Codazzi2}
    \begin{split}
        & R^{zi}_{jk}=D_{j}\theta^{zi}_{\;\;k}-D_{k}\theta^{zi}_{\;\;j}\,,\\
        & R^{ij}_{kl}=\mathcal{R}^{ij}_{kl}-\theta^{zi}_{\;\;l}\theta^{j}_{zk}+\theta^{zi}_{\;\;k}\theta^{j}_{zl}\,.
    \end{split}
\end{equation}

With the above discussion in mind, we can propose a covariant version of the renormalized gravity action
\begin{equation}
    I_\mathrm{ren}=I_\mathrm{bulk}+c_{d}\int\limits_{\partial
M}d^{d}x\,\sqrt{-h}\, n_{\sigma}B^{\sigma}_{d}\,,
\end{equation}
which clearly comes from the addition of a total derivative to the Lagrangian, i.e.,
which corresponds to the change in the Lagrangian density  $\mathcal{L}_{\mathrm{ren}}=\mathcal{L}_{\mathrm{bulk}}+c_{d}\,\nabla _{\sigma }B_{d}^{\sigma }$.

The Kounterterms are then written as a vector which, in even dimensions, take\mc{\cancel{s}} the form
\begin{eqnarray}
B^{\sigma}_{2n-1} &=&n\int\limits_{0}^{1}du\,\delta _{\mu _{1}\mu _{2}\cdots \mu _{2n}}^{\sigma \sigma_{2}\cdots \sigma
_{2n}}\,\theta _{\;\;\;\,\sigma _{2}}^{\mu _{1}\mu _{2}}\left( \frac{1}{2} \,R_{\sigma _{3}\,\sigma _{4}}^{\mu _{3}\,\mu _{4}}+(u^{2}-1)\,\theta _{\;\;\sigma _{3}}^{\lambda \mu _{3}}\theta _{\lambda \;\,\sigma _{4}}^{\mu _{4}}\right) \times \cdots  \notag \\
&&\qquad \cdots \times \left( \frac{1}{2}\,R_{\sigma _{2n-1}\,\sigma _{2n}}^{\mu_{2n-1}\,\mu _{2n}}+(u^{2}-1)\theta _{\;\;\sigma _{2n-1}}^{\lambda \,\mu_{2n-1}}\theta _{\lambda \;\,\sigma _{2n}}^{\mu _{2n}}\right)
\,. \label{CovariantKountertermeven}
\end{eqnarray}
%\sout{where} \mc{Here,} $\theta^{\mu}_{\alpha \beta}$ is \mc{a tensor} defined \sout{by} \mc{as} the difference between a dynamic Christoffel symbol $\Gamma^{\mu}_{\alpha \beta}$ and a reference, fixed connection $\bar{\Gamma}^{\mu}_{\alpha \beta}$, that is,
%\begin{equation}\label{Theta}
   % \theta^{\mu}_{\alpha \beta}=\Gamma^{\mu}_{\alpha \beta}-\bar{\Gamma}^{\mu}_{\alpha \beta}\,.
%\end{equation}
%\mc{In order to reproduce the previous result in the frame \eqref{NormalCoordinates}, the reference connection has to be chosen such that $\theta^i_{j z}=K^i_j$ is the only non vanishing component of $\theta^{\mu}_{\alpha \beta}$ on $\partial M$. For more details, see Appendix \ref{SFF}.}

For the case of odd bulk dimensions, the boundary term is covariantized in an analogous manner,
\begin{eqnarray}
B^{\sigma}_{2n} &=&n\int\limits_{0}^{1}du\int\limits_{0}^{u}ds\,\delta _{\mu _{1}\mu _{2}\cdots \mu _{2n}}^{\sigma \sigma_{2}\cdots \sigma
_{2n}}\,\theta _{\;\;\;\,\sigma _{2}}^{\mu _{1}\mu _{2}}\left( \frac{1}{2} \,R_{\sigma _{3}\,\sigma _{4}}^{\mu _{3}\,\mu _{4}}+(u^{2}-1)\,\theta _{\;\;\sigma _{3}}^{\lambda \mu _{3}}\theta _{\lambda \;\,\sigma _{4}}^{\mu _{4}}+\frac{s^{2}}{\ell_{\rm{eff}} ^{2}}\,\delta _{\sigma _{3}}^{\mu _{3}}\delta _{\sigma _{4}}^{\mu _{4}}\right)  \times \cdots \notag \\
&&\qquad \cdots \times \left( \frac{1}{2}\,R_{\sigma _{2n-1}\,\sigma _{2n}}^{\mu_{2n-1}\,\mu _{2n}}+(u^{2}-1)\theta _{\;\;\sigma _{2n-1}}^{\lambda \,\mu_{2n-1}}\theta _{\lambda \;\,\sigma _{2n}}^{\mu _{2n}}+\frac{s^{2}}{\ell_{\rm{eff}} ^{2}}\,\delta _{\sigma_{2n-1}}^{\mu _{2n-1}}\delta _{\sigma_{2n}}^{\mu _{2n}}\right)
\,.  \label{CovariantKountertermodd}
\end{eqnarray}
%
 %In the case of gravity with quadratic-curvature couplings in the bulk action, the corresponding renormalized version is given by the prescription
%\begin{equation}
    %I_{\mathrm{QCG}}^{\mathrm{ren}} =I_{\mathrm{QCG}}+c_{d}\int\limits_{\partial\mathcal{M}}\!\!d^{d}x\sqrt{-h}\,n_{\sigma}B^{\sigma}_{d}\,,
%\end{equation}
As a consequence, the prepotential $Q^{\mu \nu}$ is modified according to
\begin{equation}\label{Prepotencialmod}
   Q^{\mu \nu}_{\mathrm{ren}}(\xi)=Q^{\mu \nu}(\xi) +c_d\,\xi^{[\mu}B_{d}^{\nu]}\,\mc{,}
\end{equation}
consistent with the previous expressions for $n_{\mu} B^{\mu}_d =B_d$.

%%%%%%%%%%%%%%%%%%%%%%%%%%%%%%%%%%%%%%%%%%%%%%%%%%%%


\begin{thebibliography}{0}

\bibitem{tHooft:1974toh}
G.~'t Hooft and M.~J.~G.~Veltman,
\textit{One loop divergencies in the theory of gravitation},
Ann. Inst. H. Poincare Phys. Theor. A \textbf{20} 69 (1974).

\bibitem{Goroff:1985th}
M.~H.~Goroff and A.~Sagnotti,
\textit{The Ultraviolet Behavior of Einstein Gravity},
Nucl. Phys. B \textbf{266}, 709 (1986).

\bibitem{Stelle:1976gc}
K.S.~Stelle,
\textit{Renormalization of Higher Derivative Quantum Gravity},
Phys. Rev. D \textbf{16}, 953 (1977).

\bibitem{Grumiller:2008qz}
D.~Grumiller and N.~Johansson, 
\textit{Instability in cosmological topologically massive gravity at the chiral point},
J. High Energy Phys. \textbf{07}, 134 (2008).

\bibitem{Alishahiha:2011yb}
M.~Alishahiha and R.~Fareghbal, 
\textit{D-Dimensional Log Gravity},
Phys. Rev. D \textbf{83}, 084052 (2011).

\bibitem{Cai:2009ac}
R.-G.~Cai, Y.~Liu and Y.W.~Sun,
\textit{A Lifshitz black hole in four dimensional $R^{2}$ gravity}, 
J. High Energy Phys. \textbf{10}, 080 (2009).

\bibitem{Giribet:2018hck}
G.~Giribet, O.~Miskovic, R.~Olea and D.~Rivera-Betancour,
\textit{Energy in Higher-Derivative Gravity via Topological Regularization},
Phys. Rev. D \textbf{98}, %o.4,
044046 (2018).
[arXiv:1806.11075 [hep-th]]

\bibitem{Giribet:2020aks}
G.~Giribet, O.~Miskovic, R.~Olea and D.~Rivera-Betancour,
\textit{Topological invariants and the definition of energy in quadratic gravity theory},
Phys. Rev. D \textbf{101},  %o.6,
064046 (2020).
[arXiv:2001.09459 [hep-th]]

\bibitem{Iyer:1994ys}
V.~Iyer and R.~M.~Wald,
\textit{Some properties of Noether charge and a proposal for dynamical black hole entropy},
Phys. Rev. D \textbf{50}, 846 (1994).
[arXiv:gr-qc/9403028]

\bibitem{Aros:1999kt}
R.~Aros, M.~Contreras, R.~Olea, R. Troncoso, and J.Zanelli, \textit{Conserved charges for even dimensional asymptotically AdS gravity theories},
Phys. Rev. D \textbf{62}, 044002 (2000).

\bibitem{Kofinas:2006hr}
G.~Kofinas and R.~Olea,
\textit{Vacuum energy in Einstein-Gauss-Bonnet AdS gravity},
Phys. Rev. D \textbf{74}, 084035 (2006).
[arXiv:hep-th/0606253]


\bibitem{Maldacena:1997re}
J.~M.~Maldacena,
\textit{The Large N limit of superconformal field theories and supergravity},
Adv. Theor. Math. Phys. \textbf{2}, 231 (1998).
[arXiv:hep-th/9711200]


\bibitem{Gubser:2002tv}
S.~S.~Gubser, I.~R.~Klebanov and A.~M.~Polyakov,
\textit{A Semiclassical limit of the gauge/string correspondence},
Nucl. Phys. B \textbf{636}, 99 (2002).
[arXiv:hep-th/0204051]

\bibitem{Witten:1998qj}
E.~Witten,
\textit{Anti-de Sitter space and holography},
Adv. Theor. Math. Phys. \textbf{2}, 253 (1998).
[arXiv:hep-th/9802150]

\bibitem{Skenderis:2002wp}
K.~Skenderis,
\textit{Lecture notes on holographic renormalization},
Class. Quant. Grav. \textbf{19}, 5849 (2002).
[arXiv:hep-th/0209067]

\bibitem{Henningson:1998gx}
M.~Henningson and K.~Skenderis,
\textit{The Holographic Weyl anomaly},
J. High Energy Phys. \textbf{07}, 023 (1998).
[arXiv:hep-th/9806087]

\bibitem{Papadimitriou:2005ii}
I.~Papadimitriou and K.~Skenderis,
\textit{Thermodynamics of asymptotically locally AdS spacetimes},
J. High Energy Phys. \textbf{08}, 004 (2005).
[arXiv:hep-th/0505190]

\bibitem{Witten:2018lgb}
E.~Witten,
\textit{A note on boundary conditions in Euclidean gravity},
Rev. Math. Phys. \textbf{33}, 
%no.10, 
2140004 (2021).
[arXiv:1805.11559 [hep-th]]

\bibitem{Ashtekar:1984zz} 
A.~Ashtekar and A.~Magnon, 
\textit{Asymptotically anti-de Sitter space-times},
Class. Quant. Grav. \textbf{1}, L39 (1984).

\bibitem{Ashtekar:1999jx}
A.~Ashtekar and S.~Das,
\textit{Asymptotically Anti-de Sitter space-times: Conserved quantities},
Class. Quant. Grav. \textbf{17}, L17 (2000).
[arXiv:hep-th/9911230]

\bibitem{Olea:2005gb}
R.~Olea, 
\textit{Mass, angular momentum and thermodynamics in four-dimensional Kerr-AdS black holes},
J. High Energy Phys. \textbf{06}, 023 (2005).
[arXiv:hep-th/0504233]

\bibitem{Olea:2006vd}
R.~Olea,
\textit{Regularization of odd-dimensional AdS gravity: Kounterterms},
J. High Energy Phys. \textbf{04}, 073 (2007).
[arXiv:hep-th/0610230]

\bibitem{Miskovic:2009bm}
O.~Miskovic and R.~Olea,
\textit{Topological regularization and self-duality in four-dimensional anti-de Sitter gravity}, 
Phys. Rev. D \textbf{79}, 124020 (2009).
[arXiv:0902.2082 [hep-th]]

\bibitem{Anastasiou:2020zwc}
G.~Anastasiou, O.~Miskovic, R.~Olea and I.~Papadimitriou,
\textit{Counterterms, Kounterterms, and the variational problem in AdS gravity},
J. High Energy Phys. \textbf{08}, 061 (2020).
[arXiv:2003.06425 [hep-th]]

\bibitem{Anastasiou:2020mik}
G.~Anastasiou, I.~J.~Araya and R.~Olea,
\textit{Einstein Gravity from Conformal Gravity in 6D},
J. High Energy Phys. \textbf{01}, 134 (2021).
[arXiv:2010.15146 [hep-th]]

\bibitem{Anastasiou:2019ldc}
G.~Anastasiou, I.~J.~Araya, A.~Guijosa and R.~Olea,
\textit{Renormalized AdS gravity and holographic entanglement entropy of even-dimensional CFTs},
J. High Energy Phys. \textbf{10}, 221 (2019).
[arXiv:1908.11447 [hep-th]]

\bibitem{Boulware:1985wk}
D.~Boulware and S.~Deser, 
\textit{String generated gravity models},
Phys. Rev. Lett. \textbf{55}, 2656 (1985).

\bibitem{Bueno:2016ypa}
P.~Bueno, P.A.~Cano, V.S.~Min and M.R.~Visser,
\textit{Aspects of general higher-order gravities},
Phys. Rev. D \textbf{95}, 044010 (2017).

\bibitem{Yoel}
Yoel Parra-Cisterna, \textit{Energía en Gravedad de Curvatura Cuadrática en dimensiones impares}, B.Sc.~Thesis (2020) (in Spanish).

\bibitem{Kehagias:2015ata}
A.~Kehagias, C.~Kounnas, D.~L\"ust and A.~Riotto, 
\textit{Black hole solutions in $R^{2}$ gravity}, 
J. High Energy Phys. \textbf{05}, 143 (2015).

\bibitem{Stelle:2017bdu}
K.S.~Stelle, 
\textit{Abdus Salam and quadratic curvature gravity: Classical solutions}, 
Int. J. Mod. Phys. A \textbf{32}, 1741012 (2017).

\bibitem{Birmingham:1998nr}
D.~Birmingham,
\textit{Topological black holes in Anti-de Sitter space},
Class. Quant. Grav. \textbf{16}, 1197 (1999).
[arXiv:hep-th/9808032]

\bibitem{Svarc:2018coe}
R.~Svarc, J.~Podolsky, V.~Pravda and A.~Pravdova,
\textit{Exact black holes in quadratic gravity with any cosmological constant},
Phys. Rev. Lett. \textbf{121}, 
%no.23, 
231104 (2018).
[arXiv:1806.09516 [gr-qc]]

\bibitem{Klemm:1998kf}
D.~Klemm,
\textit{Topological black holes in Weyl conformal gravity},
Class. Quant. Grav. \textbf{15}, 3195 (1998).
[arXiv:gr-qc/9808051]


\bibitem{Ogushi:2001gp}
S.~Ogushi,
\textit{AdS Black Hole in R$^2$-gravity},
Contribution to {\it 6th Workshop on Non-Perturbative Quantum Chromodynamics}, 326 (2001). 
[arXiv:hep-th/0108209]

\bibitem{Deser:2002jk}
S.~Deser and B.~Tekin, 
\textit{Energy in generic higher curvature gravity theories},
Phys. Rev. D \textbf{67}, 084009 (2003).


\bibitem{Ghodsi:2019xrx}
A.~Ghodsi and M.~Siahvoshan,
\textit{A Holographic Study of the $a$-theorem and RG Flow in General Quadratic Curvature Gravity},
Eur. Phys. J. C \textbf{79},  820 (2019).
[arXiv:1907.03497 [hep-th]]

\bibitem{Eguchi:1980jx}
T.~Eguchi, P.~B.~Gilkey and A.~J.~Hanson,
\textit{Gravitation, Gauge Theories and Differential Geometry},
Phys. Rept. \textbf{66}, 213 (1980).

%\bibitem{Miskovic:2010ui}
%O.~Miskovic and R.~Olea,
%Conserved charges for black holes in Einstein-Gauss-Bonnet gravity coupled to nonlinear electrodynamics in AdS space,
%Phys. Rev. D \textbf{83} (2011), 024011
%[arXiv:1009.5763 [hep-th]].
%\bibitem{Anastasiou:2021tlv}
%G.~Anastasiou, I.~J.~Araya, C.~Corral and R.~Olea,
%Noether-Wald charges in six-dimensional Critical Gravity,
%JHEP \textbf{07} (2021), 156
%[arXiv:2105.02924 [hep-th]].


\end{thebibliography}
\end{document}